\documentclass[prd,aps,superscriptaddress,twocolumn,floatfix,10pt]{revtex4-2}
\RequirePackage{snapshot}
\usepackage[dvipsnames, usenames]{xcolor}
\definecolor{linkcolor}{rgb}{0.0,0.3,0.5}

\usepackage[unicode, colorlinks=true, linkcolor=linkcolor, citecolor=linkcolor, filecolor=linkcolor, urlcolor=linkcolor, linktocpage, breaklinks]{hyperref}
\usepackage[all]{hypcap}

\usepackage{amssymb,amsmath,amsthm,latexsym,mathrsfs,verbatim,
            mathtools,needspace,enumitem,etoolbox,graphicx,physics,
            microtype,afterpage,xspace,tabularx,lmodern,multirow,slashed}
\usepackage{orcidlink}
\usepackage{textcomp, gensymb}
\usepackage{flushend} 
\usepackage{comment}
\usepackage[T1]{fontenc}
\usepackage[utf8]{inputenc}
\hypersetup{colorlinks=true,citecolor=romared,linkcolor=romared,urlcolor=romared}
\usepackage{environ}
\usepackage{csvsimple,etoolbox}
\usepackage{tikz}
\usepackage{tensor}
\usepackage{placeins} %
\usepackage{longtable}

\definecolor{romared}{RGB}{142,0,28}

\newcommand{\be}{\begin{equation}}
\newcommand{\ee}{\end{equation}}

\def\be{\begin{equation}}
\def\ee{\end{equation}}
\newcommand{\beq}{\begin{eqnarray}}
\newcommand{\eeq}{\end{eqnarray}}

\usepackage{makecell}
\usepackage{soul}

\newcolumntype{Y}{>{\centering\arraybackslash}X}

\usepackage[normalem]{ulem}
\usepackage{adjustbox}

\begin{document}

\title{Twist-free axisymmetric critical collapse of a complex scalar field}

\author{Krinio Marouda\orcidlink{0000-0003-1030-8853}} 
\affiliation{CENTRA, Departamento de F\'{\i}sica, Instituto Superior T\'ecnico -- IST, Universidade de Lisboa -- UL, Avenida Rovisco Pais 1, 1049-001 Lisboa, Portugal}

\author{Daniela Cors\orcidlink{0000-0002-0520-2600}} 
\affiliation{Department of Applied Mathematics and Theoretical Physics, Centre for Mathematical Sciences, University of Cambridge, Wilberforce Road, Cambridge CB3 0WA, United Kingdom}

\author{Hannes R. R\"uter\orcidlink{0000-0002-3442-5360}} 
\affiliation{CENTRA, Departamento de F\'{\i}sica, Instituto Superior T\'ecnico -- IST, Universidade de Lisboa -- UL, Avenida Rovisco Pais 1, 1049-001 Lisboa, Portugal}

\author{Florian Atteneder\orcidlink{0000-0002-6617-5482}} 
\affiliation{Friedrich-Schiller-Universit\"at, Jena, 07743 Jena, Germany}

\author{David Hilditch\orcidlink{0000-0001-9960-5293}} 
\affiliation{CENTRA, Departamento de F\'{\i}sica, Instituto Superior T\'ecnico -- IST, Universidade de Lisboa -- UL, Avenida Rovisco Pais 1, 1049-001 Lisboa, Portugal}

\date{\today}

\begin{abstract}
It has been three decades since the groundbreaking work by Choptuik on critical phenomena in gravitational collapse involving a real massless scalar field minimally coupled to general relativity in spherical symmetry. These phenomena are characterized by the emergence of surprising structure in solution space, namely the appearance of universal power-laws and periodicities near the threshold of collapse, and a universal discretely self-similar solution at the threshold itself. This seminal work spurred a comprehensive investigation of extreme spherical spacetimes in numerical relativity, with analogous results for numerous matter models. Recent research suggests that the generalization to less symmetric scenarios is subtle. In twist-free axisymmetric vacuum collapse for instance, numerical evidence suggests a breakdown of universality of solutions at the threshold of collapse. In this study, we explore gravitational collapse involving a massless complex scalar field minimally coupled to general relativity. We employ the pseudospectral code \textsc{bamps} to investigate a neighborhood of the spherically symmetric critical solution in phase space, focusing on aspherical departures from it. First, working in explicit spherical symmetry, we find strong evidence that the spacetime metric of the spherical critical solution of the complex scalar field agrees with that of the Choptuik solution. We then examine universality of the behavior of solutions near the threshold of collapse as the departure from spherical symmetry increases, comparing with recent investigations of the real scalar field. We present a series of well-tuned numerical results and document shifts of the power-law exponent and periods as a function of the degree of asphericity of the initial data. At sufficiently high asphericities we find that the center of collapse bifurcates, on the symmetry axis, but away from the origin. Finally we look for and evaluate evidence that in the highly aspherical setting the collapse is driven by gravitational waves.
\end{abstract}
 
\maketitle

\section{Introduction}
\label{sec:introduction}

One of the major breakthroughs in general relativity (GR) from the previous century was the discovery of critical phenomena in gravitational collapse. Choptuik~\cite{choptuik1993universality}, in 1993, studied a real spherically symmetric self-gravitating scalar field in numerical relativity (NR), unveiling interesting properties that emerge at the threshold of collapse, such as universality, discrete self-similarity (DSS), and the power-law scaling of the final black hole mass against the distance in phase space from the critical solution. In the subcritical sector of solution space in which no black hole forms by definition, this last property manifests as a power law for curvature invariants, such as the Ricci scalar or the Kretschmann invariant. The aforementioned properties constitute the standard picture of what is usually referred to as critical phenomena in gravitational collapse, a term coined due to the resemblance to phase transitions in various fields of physics. This analogy is beautifully explained in the living review on the topic by Gundlach and Mart\'in-Garc\'ia~\cite{gundlach2007critical}. For the scalar field model in spherical symmetry, it was shown empirically that the phase space of initial data is separated by a co-dimension one hypersurface into two regions of either collapse to a black hole or dispersion, as an end state of the evolution. From the dynamical systems point of view this has a nice interpretation, in which the single unstable mode of the critical solution is characterised by a Lyapunov exponent that is the inverse of the aformentioned power-law scaling exponent. In numerical calculations, to approach this co-dimension one space, one typically adjusts a single parameter in the initial data, such as the amplitude or the width of a Gaussian initial pulse, keeping all other parameters fixed. Subsequently, one performs a search for that special value of the parameter that lies at the threshold of collapse, with usually 12-15 decimal places accuracy that is provided by double precision arithmetic. This critical parameter lies on the separator hypersurface in phase space, close to which near-zero mass black holes are being created and, therefore, in the limit of infinite tuning, it probes a naked singularity. Thus, a huge interest has been raised in studying these types of systems, using different matter models and more generic symmetry assumptions, in order to robustly examine universality of the critical solution.

After thirty years of progress in the field, there is very strong evidence that the Choptuik critical solution for the scalar field model is universal in spherical symmetry. It has been systematically validated in the `numerical laboratory' of various independent gravitational collapse codes, with various formulations of GR and discretizations~\cite{gundlach2007critical}. Nevertheless, the extent to which this universality carries over with initial data containing arbitrary non-spherical perturbations is not absolutely clear. In fact, numerical work shows that the Choptuik solution is not universal under a more relaxed symmetry hypothesis. As we discuss in detail momentarily, evidence suggests that the threshold of collapse becomes more subtle as symmetry is discarded. 

Developments in twist-free axisymmetric vacuum collapse for instance~\cite{alcubierre2000gravitational, rinne2008constrained, sorkin2010axisymmetric, hilditch2013collapse, Hilditch:2017dnw, khirnov2018slicing, Ledvinka:2021rve, fernandez2022evolution} indicate behavior that deviates significantly from the standard picture. More precisely, in the collapse of gravitational waves, universality and DSS are being questioned following numerous unsuccessful attempts to reproduce earlier results by Abrahams and Evans~\cite{abrahams1993critical}. The terminology twist-free is inherited from the Geroch-decomposition~\cite{Geroch:1970nt}, which is beautifully explained in Rinne's thesis~\cite{rinne2013axisymmetric}. Physically, non-vanishing angular momentum implies non-vanishing twist, so we are concerned with a class of axisymmetric spacetimes that contains all those with vanishing angular momentum. In practice the vacuum works just mentioned, plus our present study, also assume vanishing angular momentum. Interestingly, three independent research groups~\cite{baumgarte2023critical} have recently reached a consensus on the fact that, at least with the more limited degree of tuning currently possible in this setting, some signs of universality do not occur in the collapse of gravitational waves. The scaling exponents do not appear to be universal. The locations of centers of collapse turn out to be family dependent. Moreover, none of these groups found evidence indicating the presence of exact DSS in the collapse of gravitational waves.

In fact, the vacuum scenario differs qualitatively from  Choptuik's classical setup, because it does not admit a non-trivial dynamical spherical limit in 3+1 dimensions. This makes the understanding of how it relates to the standard picture of critical collapse tricky. The urge to interpret subtleties appearing in the axisymmetric category leads naturally to the thought of studying models that admit dynamical black hole formation in spherical symmetry and with arbitrarily small axisymmetric deformations. For the real scalar field such configurations have been studied in a number of works. Nonlinear numerical evolutions were presented by Choptuik et al.~\cite{choptuik2003critical}, Baumgarte~\cite{baumgarte2018aspherical} and Gundlach et al.~\cite{Gundlach:2024eds} in axisymmetry and by Healy and Laguna in~\cite{Healy:2013xia}, and Deppe et al.~\cite{deppe2019critical}, in full 3d. 
In~\cite{Clough:2016jmh}, Clough and Lim study the critical collapse of a real scalar field, but with a nonlinear potential. 
Unsurprisingly all of the studies in full 3d without symmetry achieve less tuning to the threshold of collapse. Again in full 3d, but working perturbatively to linear order, Mart\'in-Garc\'ia \& Gundlach~\cite{martin1999all} found that all nonspherical mode solutions on the Choptuik spacetime decay. In~\cite{baumgarte2018aspherical} an interpretation was offered that with large enough departures from sphericity the slowest of these decay rates actually flips sign such that aspherical modes grow. Thus, consistent with the results of~\cite{choptuik2003critical}, Baumgarte finds that this growth leads to a bifurcation of the data in which two centers of collapse appear. 

Physically speaking, generic aspherical spacetimes consist of a combination of gravitational wave and matter content. There is thus a competition between the two as a possible route to black hole formation. In axisymmetric vacuum collapse, such a bifurcation of local maxima of the curvature is also observed. An obvious interpretation of bifurcation in scalar field collapse is therefore that, in very aspherical spacetimes near the threshold of collapse it is the gravitational wave content that drives the collapse, with all of the subtleties around criticality discussed above then inherited by the composite system. But concrete evidence in favor of this explanation is lacking.

In the light of the above, in this work we explore aspherical spacetimes obtained with a different matter model, namely the complex scalar field minimally coupled to GR. In the gravitational critical collapse literature, there exist two references regarding the study of the complex scalar field without a non-linear potential. The first is the massless case with angular momentum in axisymmetry, studied in~\cite{choptuik2004critical}, where the authors explicitly state that the critical solution they find is far away in solution space from the spherical case. Our focus is instead on a region of solution space including small deviations from spherical symmetry, from perturbatively small to very large ones. The results of~\cite{choptuik2004critical} suggest that the DSS threshold solution arrived at by tuning in the manner described above is universal ``within the imposed symmetry class''. Interestingly, this solution appears with different values of the echoing period and the scaling exponent than those found for the real massless scalar field. Our aim is instead to study aspherical deformations of the spherical critical solution. Thus our class of initial data, which is axisymmetric and twist-free and so has vanishing angular momentum, lies outside the symmetry class of~\cite{choptuik2004critical}. Presently we do not consider the transition into that class. We also leave the case of massive complex scalar fields to future work. Those have been studied already in spherical symmetry~\cite{jimenez2022critical}, the second complex scalar field critical collapse study referred to above, but investigations of aspherical setups are still missing.

In sections~\ref{sec:massless_complex_scalar_field}
and~\ref{sec:formulation_and_numerical_methods} we give an introduction to the system under investigation, an overview of the formulation of the equations and technical details of the code. In section~\ref{sec:phase_space_search} we discuss our diagnostics of collapse. In section~\ref{sec:numerical_results} we present numerical results for various types of initial data. The first set of results reproduces the well understood spherically symmetric picture for our model, while the following set attempts to elucidate the behavior in axisymmetry. Our initial data for aspherical deformations include several families that incorporate a linear combination of an~$l=0$ and an~$l=2$ spherical harmonic mode, which allows us to parametrize the degree of asphericity and establish a direct comparison with the spherical configuration, in a similar manner to that of~\cite{choptuik2003critical,baumgarte2018aspherical}. We report the development of a disjoint pair of centers of collapse in our simulations for some axisymmetric families for the first time in the collapse of a complex scalar field. Additionally, we observe drifts in the exponents of the scaling laws and in the echoing period of the DSS-echoing periods as a function of the amount of asphericity. These results are in qualitative agreement with those of~\cite{choptuik2003critical,baumgarte2018aspherical} for the real scalar field. We then examine the extent to which gravitational wave content increases with asphericity. Section~\ref{sec:conclusions} contains our conclusions.

\section{Massless complex scalar field}
\label{sec:massless_complex_scalar_field}

The case of a massless complex scalar field minimally coupled to general relativity can, in fact, be viewed as the competition between two real scalar fields, the real and imaginary part that constitute the complex field, which interact only through their interplay with the gravitational field. The corresponding action of this system, written in geometric units $G=c=1$, is 
\begin{align}
    S=\int \mathrm{d}^4 x \sqrt{-g} \left(\frac{R}{16\pi}-\frac{1}{2}\nabla_a \psi \nabla^a \psi^\dagger \right)\,
\end{align}
where $R$ is the Ricci scalar, $g$ is the determinant of the metric, $\psi$ is the complex scalar field and $\psi^\dagger$ its complex conjugate. Here and throughout, Latin indices starting from~$a$ denote spacetime components, whereas indices starting from~$i$ denote the spatial components (when working in standard 3+1 coordinates). The equations of motion of this theory obtained via the variational principle are simply Einstein's equations coupled to the massless Klein-Gordon equation, namely
\begin{align}
    G_{ab} &=R_{ab}-\frac{1}{2}g_{ab}R =8\pi T_{ab}\,,\label{eqn:EEs}\\
    \square \psi &= g^{ab} \nabla_a \nabla_b \psi =0\,.
\end{align}
The stress energy tensor is given by
\begin{align}
    T_{ab}=\nabla_{\left(a\right.} \psi  \nabla_{\left.b\right)}\psi^\dagger -\frac{1}{2}g_{ab} \nabla^c \psi \nabla_c \psi^\dagger\,.
\end{align}
Writing $\psi=\psi^{\textrm{Re}}+i \psi^{\textrm{Im}}$, the stress energy tensor can be decomposed as $T_{\mu\nu}=T_{\mu\nu}^\textrm{Re}+T_{\mu\nu}^{\textrm{Im}}$, with each of these corresponding to the standard expression for the stress-energy tensor of a real massless scalar field, evaluated on the real and imaginary part of the field, respectively. Note that since there is no direct interaction between the real and imaginary parts of the complex field, so each individually satisfies the massless Klein-Gordon equation. Such a split will not, in general, be possible in the presence of a potential, but does make the interpretation of the dynamics straightforward in the massless case. Due to the fact that the real and imaginary parts of the complex scalar field follow exactly the same equation as the real scalar field itself, interacting only through their minimal coupling to gravity, we expect a qualitative behavior similar to the real scalar field case. In this way, one could, in principle, consider the massless complex scalar field case as a supplementary study of the real scalar field case, albeit with a bigger solution space and correspondingly richer dynamics due to the global internal~$U(1)$ symmetry of the complex field.

\section{Formulation and numerical methods}
\label{sec:formulation_and_numerical_methods}

The evolution tool we use for this work is \textsc{bamps}, a pseudospectral code for the time development of first order symmetric hyperbolic systems. In this section we give a brief overview of the continuum equations solved, our grid setup, $hp$ mesh-refinement strategy and our improved constraint damping parameters. Full details of these aspects of the code can be found in the technical references~\cite{Bruegmann:2011zj,hilditch2016pseudospectral,renkhoff2023adaptive,cors2023formulation}.

\subsection{Formulation of General Relativity}

\subsubsection{Geometry}

In our continuum formulation of GR we follow the approach of~\cite{Lindblom:2005qh}, first imposing generalized harmonic gauge (GHG) in the well-known fashion, see for instance~\cite{Pretorius:2004jg}, before introducing all first derivatives of metric components as evolved quantities. Sensible choices in this reduction result in a first order symmetric hyperbolic formulation of GR. The formulation is bolstered by a damping scheme for the harmonic constraints essentially lifted from~\cite{Gundlach:2005eh} along with improved constraint damping parameters for the reduction and harmonic constraints introduced in~\cite{cors2023formulation}. 

More specifically, the first-order reduction is accomplished by introducing the time reduction variable $\Pi_{ab}=-n^c\partial_c g_{ab}$, with~$n^a$ denoting as usual the future pointing unit normal to level-sets of constant time~$t$, and the spatial reduction variable $\Phi_{iab}=\partial_i g_{ab}$. The associated reduction constraint for the latter is $C_{iab}:=\partial_i g_{ab}-\Phi_{iab}$. The evolved variables are $g_{ab}$, $\Pi_{ab}$ and $\Phi_{iab}$ and the evolution equations as they appear also in~\cite{cors2023formulation} are
\begin{align}
    \partial_t g_{ab}&= \beta^i \partial_i g_{ab} - \alpha \Pi_{ab} + \gamma_1 \beta^i C_{iab}\,, \\
    \partial_t \Pi_{ab}&= \beta^i \partial_i \Pi_{ab}-\alpha \gamma^{ij} \partial_i \Phi_{jab}+\gamma_1\gamma_2\beta^i C_{iab} \\
    &+ 2\alpha g^{cd} \left(\gamma^{ij} \Phi_{ica}\Phi_{jdb}-\Pi_{ca}\Pi_{db}-g^{ef}\Gamma_{ace}\Gamma_{bdf}\right) \nonumber \\
    &- 2\alpha \left( \nabla_{\left(a\right.} H_{\left. b\right)} +\gamma_4 \Gamma^c_{\, ab} C_c -\frac{1}{2}\gamma_5 g_{ab}\Gamma^c C_c\right) \nonumber\\
    &-\frac{1}{2}\alpha n^c n^d \Pi_{cd}\Pi_{ab}-\alpha n^c \gamma^{ij}\Pi_{ci}\Phi_{jab} \nonumber\\
    &+ \alpha \gamma_0 \left(2\delta^c_{\left(a\right.}n_{\left.b\right)}-g_{ab}n^c\right)C_c\nonumber \\
    &-16\pi\alpha\left(T_{ab}-\frac{1}{2}g_{ab}T^c{}_{c}\right)\nonumber\,,\\
    \partial_t \Phi_{iab}& = \beta^j\partial_j \Phi_{iab} -\alpha \partial_i \Pi_{ab}+ \gamma_2 \alpha C_{iab} \\
    &+\frac{1}{2}\alpha n^c n^d \Phi_{icd}\Pi_{ab}+\alpha\gamma^{jk}n^c\Phi_{ijc}\Phi_{kab}\nonumber\,.
\end{align}
where $\alpha \gamma_0 =\alpha \gamma_2 =  2$, $\gamma_1 = -1$, 
and $\gamma_4=\gamma_5=\frac{1}{2}$. We define the lapse~$\alpha$, shift~$\beta^i$ and spatial metric~$\gamma_{ij}$ according to the standard notation in NR. Following the suggestion presented in~\cite{cors2023formulation}, we have adapted our damping parameters $\gamma_0$ and $\gamma_2$ to collapse spacetimes by adding the lapse in their definition. At the outer boundary we employ constraint-preserving, radiation-controlling boundary conditions. These conditions work by setting the incoming gravitational wave degrees of freedom, captured by the Weyl scalar~$\Psi_0$, to vanish, and likewise doing the same for all incoming characteristic variables of the constraint subsystem. See~\cite{Rinne:2006vv,Ruiz:2007hg,hilditch2016pseudospectral} for details. 

\subsubsection{Matter}

For the matter content of the spacetime we introduce the time and space reduction variables $\Pi = n^a\partial_a \psi$, $\Phi_i=\partial_i \psi$ and the spatial reduction constraint is defined as $S_i :=\partial_i \psi-\Phi_i$. The evolution equations for the real and the imaginary parts of the scalar field take the following form accordingly
\begin{align}
    \partial_t \psi&=\alpha \Pi + \beta^i \Phi_i\,, \\
    \partial_t \Phi_i &=\Pi \partial_i\alpha+ \alpha\partial_i \Pi+\sigma\alpha S_i+\Phi_j\partial_i\beta^j+\beta^j\partial_j \Phi_i\,,\\
    \partial_t \Pi&=  \beta^i \partial_i \Pi +\alpha\Pi K +\sigma \beta^i S_i \nonumber\\
    &+ \gamma^{ij}\big(\Phi_j\partial_i\alpha+\alpha\partial_i \Phi_j -\alpha {}^{(3)}\Gamma^k{}_{ij}\Phi_k\big)\,.
\end{align}
The evolved variables are $\Phi^\textrm{Re}_i$, $\Phi^\textrm{Im}_i$,
$\Pi^\textrm{Re}$, $\Pi^\textrm{Im}$, $ \psi^\textrm{Re}$ and 
$\psi^\textrm{Im}$. Here, $\partial_i \alpha$ is determined from the evolved variables as~$\partial_i \alpha= -\frac{1}{2}\alpha n^a n^b \Phi_{iab}$. The trace of the extrinsic curvature~$K$ and the spatial Christoffel symbols~${}^{(3)}\Gamma^k{}_{ij}$ are similarly evaluated by assuming that the reduction constraints are satisfied and taking appropriate combinations of the reduction variables. In practice, our implementation is that used for binary boson star evolutions in~\cite{Atteneder:2023pge} but with the potential suppressed. This was based in turn on our implementation of the real scalar field as used in~\cite{Bhattacharyya:2021dti,cors2023formulation}, and so has already been well tested in the past. The damping parameter for the scalar field constraint is selected to be $\alpha \sigma=2$, similar to our treatment of $\gamma_2$. At the outer boundary we employ reduction constraint preserving boundary conditions together with Sommerfeld-like conditions on the incoming physical characteristic variables.

\subsubsection{Initial Data}
\label{subsubsection:ID}

We construct conformally flat initial data, $\bar \gamma_{ij} := \gamma_{ij}/\psi_\mathrm{conf}^4 = \delta_{ij}$ and $\partial_t \bar \gamma_{ij} = 0$, with maximal slicing, $K=0$ and $\partial_t K=0$, using the extended conformal thin-sandwich~(XCTS) formulation~\cite{PfeYor03,BauSha10,Tic17} of the constraint equations. At the outer boundary we impose Robin boundary conditions on $\psi_\mathrm{conf}$, $\alpha$, $\beta^i$ compatible with a $1/r$ decay towards the flat space values. The initial data for the complex scalar field is specified in Sec.~\ref{sec:numerical_results}.

The XCTS equations form a system of elliptic partial differential equations, that we solve by means of the hyperbolic relaxation method implemented in \textsc{bamps}~\cite{Ruter:2017iph}.

\subsection{Gauge}

In the GHG formulation, the coordinates are set to follow wave equations with sources, with the latter being called the gauge source functions, which constitute the remaining gauge freedom of the system. To straightforwardly maintain well-posedness at the partial differential equation level, those functions are allowed to depend on the metric, but not on its derivatives, $H_a=H_a(g)$. A good choice of those functions may lead to coordinates capable of resolving the delicate properties or symmetries of the physical system under investigation. In the particular case of critical phenomena, the primary difficulty of any numerical study is the extreme resolution required at the center of collapse due to physics unfolding at ever smaller scales. By definition a DSS spacetime is one which admits highly non-unique coordinates~$T,x^i$ in which the metric takes the form,
\begin{align}
g_{ab}=e^{-2T}\tilde{g}_{ab}(T,x^i)\,,
\end{align}
with~$\tilde{g}_{ab}(T,x^i)$, a conformal metric~$\Delta$-periodic in~$T$, commonly referred to as slow-time. The physical metric therefore has a curvature singularity in the limit~$T\to\infty$ at any~$x^i$ where~$\tilde{g}_{ab}$ is not flat~\cite{gundlach2007critical}. Thus, if we consider a family of metrics close in solution space to a DSS limit, $\Delta$ will manifest as the period between repeated features, also called echoes, in a suitably adapted time coordinate. The change in physical scale between one echo and the next goes like~$e^\Delta$. For the spherical real massless scalar field it turns out to be~$e^\Delta\approx 31$ in the strong-field region for a spacetime tuned to the threshold of collapse within a generic smooth one parameter family of initial data. Consequently, unless coordinates that dynamically adapt to the presence of the symmetry are chosen, in practice it is expected that numerical resolution must be increased from one echo to the next to account for this and thus to maintain reasonable errors. Several coordinate choices that do the job are known in spherical symmetry; for instance single/double-null. Since the solution space for the complex scalar field contains that of the real case, one would expect similar values also with the model we treat here. Unfortunately there are no known gauge source functions for GHG that guarantee such a dynamical adaptation to DSS in a general setting. To allow for the possibility of dynamically constructing DSS-adapted coordinates during the DSS phase, we make a choice for the gauge source functions that satisfies necessary conditions for the coordinates to respect such a symmetry, the DSS-compatible gauge choice introduced in~\cite{cors2023formulation}. (In practice for the real scalar field the resulting coordinates turn out not to adapt perfectly to the DSS of the Choptuik solution, but it was found that they behave at least as well as, or better than, alternative suggestions in the literature, so we use them unmodified here. Considering all this, and the fact that the spacetime location of the~$T\to\infty$ events are not known a priori, we turn to the generic adaptive mesh algorithm summarized below).

In general, the GHG constraints read
\begin{align}
    C_a=\Gamma_a+H_a=0\,,
\end{align}
with the contracted Christoffel symbol defined as usual by tracing the Christoffel symbols with the inverse metric~$\Gamma^a=g^{bc}\Gamma^a{}_{bc}$. This is defined uniquely in a first order reduction only up to addition of reduction constraints. By convention, within the reduction we therefore use the reduction variables to remove all explicit derivatives from the GHG constraints. The translation of the contracted Christoffel symbol into standard~$3+1$ notation gives the following evolution equations for the lapse and shift,
\begin{align}
    d_t \alpha &= -\alpha^2\left(n^a H_a +K\right)\,,\\
    d_t \beta^i &=  \alpha^2\left(H^i+ {}^{(3)}\!\Gamma^i-\partial^i \ln{\alpha}\right)\,,
\end{align}
with the spatial contracted Christoffel symbols defined as~${}^{(3)}\!\Gamma^i=\gamma^{jk}{}^{(3)}\!\Gamma^i{}_{jk}$, and where $d_t=\partial_t -\beta^k \partial_k$. In our case, the DSS compatible gauge that we are using is
\begin{align}
    H_a&= \eta_L \alpha^{-1}\ln \left(\alpha^{-3}\sqrt{\gamma}\right)n_a-\eta_S\gamma_{ai}\beta^i\alpha^{-2}\,,\label{eqn:DSS_compatible}\\
    d_t\alpha&=\eta_L\alpha\ln\left(\alpha^{-3}\sqrt{\gamma}\right)-\alpha^2 K\,, \\
    d_t\beta^i&=-\eta_S\beta^i +\alpha^2\left( {}^{(3)}\!\Gamma^i-\partial^i \ln \alpha\right)\,,
\end{align}
with $\eta_L=4$ and $\eta_S=6$. Here, ``compatibility with DSS'' means that it is possible that the gauge source functions satisfy 
\begin{align}
    H_a \left(T+\Delta,x^i\right)=H_a \left(T,x^i\right)\,,
\end{align}
during the evolution, a characteristic that is not shared with other commonly used gauge choices. Gauge sources that violate this condition can never result in DSS-adapted coordinates. 

\subsection{Grid setup}

In \textsc{bamps}, the computational domain is divided into atomic, cubic, grids, each one of those solving for its very own Initial Boundary Value Problem (IBVP). The basic communication of data between these grids is performed using the penalty method as explained in~\cite{hilditch2016pseudospectral}. When neighboring grids have differing resolution, an additional interpolation is required for this communication~\cite{renkhoff2023adaptive}. Within each grid, the numerical solutions are approximated by a nodal pseudospectral expansion based on Gauss-Lobatto-Chebyshev points in each dimension. This provides an efficient approximation of the spatial derivatives throughout the evolution. The discretization and evolution in time is performed using the method of lines through a 4th order Runge-Kutta (RK4) scheme.

The numerical domain that approximates the spatial slices in \textsc{bamps} is composed of patches of these local grids, which in their turn form three main regions, a cube in the strong field region, a spherical outer shell and a cubed-sphere shell in-between for matching purposes, see Figure~$1$ in~\cite{hilditch2016pseudospectral}. The choices we make for the size of the central box, the transition shell and the outer sphere radius are adapted to each initial data set, so that the strong field region is inside the central box. For further details regarding the grid configurations, see Table~\ref{tbl:grid_parameters}.

\begin{table}[t!]
 \centering
 \begin{ruledtabular}
 \begin{tabular}{lllll}
      Parameter & Setup 1 & Setup 2 & Setup 3 & Setup 4 \\
      \hline
      \texttt{grid.cube.max}                   & 3  &  3    &  4    & 14  \\
      \texttt{grid.sub.xyz}                   & 12  &  6   &  16   & 32 \\
      \texttt{grid.cubedsphere.max.x}  & 14  &  14   &  15   & 30 \\
      \texttt{grid.cubedsphere.sub.x}  & 6   &  4  &  8    & 8 \\
      \texttt{grid.sphere.max.x}       & 30  &  30   &  30   & 56 \\
      \texttt{grid.sphere.sub.x}       & 5   &   5  &  6    & 13 \\
      \texttt{grid.dtfactor}           &0.25  & 0.25     &  0.25 & 0.25 \\
      \texttt{grid.cartoon}            & x     & x        &  xz   & xz\\
      \texttt{grid.reflect}            & z     & z        &  z    & z \\
      \texttt{grid.n.xyz}              & [5,11]& [7,19]  & [5,13] & [5,13] \\
  \end{tabular}
  \end{ruledtabular}
 \label{tbl:grid_parameters}
 \caption{
   Summary of the grid setup used for all the families presented in this work. Setup 1: Family III from Table~\ref{tbl:sph_fam_realim}, Family I from Table~\ref{tbl:axi_families} , Setup 2: Families I and II from Table~\ref{tbl:sph_fam_realim}, Setup 3: Families II and III from Table~\ref{tbl:axi_families}, Setup 4: Family IV from Table~\ref{tbl:axi_families}.
   Parameter description:
   \texttt{grid.cube.max}: 1/2 side length of inner cube,
   \texttt{grid.sub.xyz}: number of subdivisions in inner cube,
   \texttt{grid.cubedsphere.max.x}: outer radius of cube-to-sphere patch,
   \texttt{grid.cubedsphere.sub.x}: number of radial subdivisions in cube-to-sphere patch,
   \texttt{grid.sphere.max.x}: outer radius of sphere patch,
   \texttt{grid.sphere.sub.x}: number of radial subdivisions in sphere patch,
   \texttt{grid.dtfactor}: CFL factor,
   \texttt{grid.cartoon}: the remaining dimensions after applying the cartoon method,
   \texttt{grid.reflect}: reflection symmetry across $z=0$ plane,
   \texttt{grid.n.xyz}: number of points per grid and per dimension, which increase in steps of two.
 }
\end{table}

\subsection{\texorpdfstring{$hp$}{hp}-refinement}

As motivated above, our spacetimes of interest are extreme in terms of their need for physical resolution in the strong field region. Since there are no known gauge source functions for GHG that result in DSS-adapted coordinates in any generality, we rely on mesh-refinement to achieve the required resolution. This is done in two complementary ways, namely by decreasing the size of individual grids ($h$-refinement) or by increasing the polynomial order for the spectral decomposition within a single grid, which translates to augmenting the number of collocation points in a local grid ($p$-refinement). In this work, we use $hp$-adaptive mesh refinement as presented in~\cite{renkhoff2023adaptive}. For $h$-refinement the code uses a criterion related to the smoothness of the solution, while for $p$-refinement an error estimate based on the truncation order of the spectral expansion per grid is used.

We employ a maximum of~$15$ levels of~$h$-refinement, where we halve the grid each time that we refine, and a maximum of~$31$ points per dimension per grid with a step of two in $p$-refinement. Refinement, or de-refinement, in~$h$ or~$p$ is triggered whenever the associated indicator yields a value~$\epsilon$ which falls outside a parameter range~$[\epsilon_{\textrm{min}}, \epsilon_{\textrm{max}}]$. The smoothness indicators that we used for $h$-refinement are~$[0.0005,0.001]$ for the spherical runs and~$[0.005,0.01]$ for the axisymmetric ones. The refinement criterion for $p$-refinement is a target error between~$[10^{-12},10^{-10}]$ for all simulations. Following~\cite{renkhoff2023adaptive}, we monitor the raw metric and scalar field variables with the indicators to trigger refinement. The time evolution method employs Message Passing Interface (MPI) parallelization to distribute work evenly over distributed resources throughout an evolution. Our initial data construction is made using the hyperbolic relaxation method discussed above in section~\ref{subsubsection:ID}. Our spherically symmetric evolutions were performed on a desktop machine. The aspherical runs were done using 768 CPU cores on SuperMUC-NG.

\section{Phase space search}
\label{sec:phase_space_search}

The existence of a black hole in an asymptotically flat spacetime solution is associated to the existence of an event horizon, a global property of the spacetime, which requires a global treatment of all the evolution data in order to be located. Another notion of a horizon, which is \textit{local in time}, is that of an apparent horizon.
It is defined as the outermost ``marginally outer trapped surface'' (MOTS) living on a Cauchy slice, $\Sigma$, of the spacetime. A MOTS is a compact 2-dimensional submanifold of~$\Sigma$, $S \subseteq \Sigma$, with the property that the outgoing null geodesics emanating from it have a zero expansion $\Theta$. The existence of a trapped surface, alongside energy condition assumptions and global hyperbolicity, is well-known to lead to singular, in the sense of geodesically incomplete, spacetimes~\cite{Penrose:1964wq}. 
If the cosmic censorship conjecture holds, then an apparent horizon is located inside the event horizon and for stationary spacetimes coincides with it.
It therefore serves as a sensible approximation for the event horizon. Furthermore, because it is a \textit{local in time}, but \textit{non-local in space} notion of a horizon, it is a much more efficient diagnostic of collapse from a numerical perspective, since it can be calculated accurately during the evolution.

The expansion of null geodesics for an arbitrary 2d surface with normal vector $s^i$ is 
\begin{align}
    \Theta \equiv D_i s^i + K_{ij}s^i s^j-K \, ,
\end{align}
where $K_{ij}$ is the extrinsic curvature, $K$ its trace, and $D_i$ is the induced spatial covariant derivative. The expansion of outgoing null geodesics is everywhere zero on a MOTS. In the spherically symmetric case, this condition reduces to an algebraic expression, which in adapted coordinates takes the form
\begin{align}
    \Theta=\frac{\partial_r g_{\theta\theta}}{g_{\theta\theta}\sqrt{g_{rr}}}- 2\frac{K_{\theta\theta}}{g_{\theta\theta}}=0 \, .
\end{align} 
Hence, the task of locating an apparent horizon simply transforms to a root finding process and the threshold search is performed via a binary search algorithm. In the case of axisymmetry, however, one has to solve a differential equation instead. The numerical tool we are using is \textsc{AHloc3d}~\cite{ahloc3d}, which uses a flow method starting from a large arbitrary 2d surface, in combination with a Newton-Raphson method in the end. A shortcoming of this method is that for a horizon that is not star-shaped our tool is unable to locate it, since this assumption is used explicitly in the search algorithm. Moreover, it is a post-processing tool that requires expensive output, both in terms of time and storage, from our evolution code. Our search strategy at each bisection level is thus to perform a series of runs, with some of them obviously dispersing to infinity (subcritical) and some others blowing up (potentially supercritical). Subsequently, we reproduce the ones that blow up, starting from check-pointed data late in the evolution, but now additionally writing output that is destined as an input for \textsc{AHloc3d} which then provides a proper classification. 

In some of our apparent horizon searches in highly tuned spacetimes of families III and IV in Table~\ref{tbl:axi_families} below, the flow method of the finder seemed to work up to outgoing null expansions of order~$\sim 0.0001$, but then the Newton step was unable to detect a horizon. In these cases of seemingly supercritical runs where our horizon finder struggles to detect a horizon, our diagnostic of collapse becomes the observation of the behavior of the curvature with respect to the constraint violation as it tends to blow up. In case it is an abrupt blow-up we would understand that it is probably a result of constraint violation, in which case the spacetime is considered unclassified. On the other hand, if the blow-up of the curvature is smooth, while the constraint violation remains~$\sim 5$ or more orders of magnitude lower than any non-vanishing curvature scalar throughout the evolution, including the phase close to the blowup, then we classify the relevant spacetime as a collapse. Nevertheless, these limitations of our diagnostic tool do not affect the tuning from the subcritical side, since dispersion is always classified unambiguously, but they might affect the accuracy of our estimation of the critical amplitude.

We would prefer to have a more reliable apparent horizon finder, but would have to generalize our present algorithm to allow non star-shaped horizons to do so. For a detailed review of horizon searches in NR see~\cite{Thornburg:2006zb}. For examples of non-star shaped horizon searches see~\cite{Pook-Kolb:2018igu,Poo20}. For more recent reviews in the form of PhD theses, see~\cite{Chu12, Poo20}. Nevertheless, a better apparent horizon approach on the supercritical side would essentially result in a smaller value in our estimation of the threshold amplitude. Looking ahead, this would affect only figures across the phase space, as opposed to those from a fixed spacetime, exaggerating the results that we report below.

\section{Numerical Results}
\label{sec:numerical_results}

\subsection{Spherical Symmetry}

We commence our numerical work in spherical symmetry, which allows us to gain an initial understanding of the underlying physics of the complex scalar field matter economically whilst recycling the computational setup used in~\cite{cors2023formulation} directly. Although we are not aware of any published work in which this precise setup is treated, we anticipate that the behavior of the complex scalar field will closely resemble that of the real scalar field. This expectation is rooted in the facts that, as previously explained, both the real and imaginary components of the complex scalar field satisfy identical evolution equations and conservation laws as the real scalar field, and in the numerical work of~\cite{jimenez2022critical} it was shown that the massive complex field behaves much like the massless real scalar field with initial data of a certain type (in which the radial extent of the data is less than the Compton wavelength~\cite{Brady:1997fj}). Starting in spherical symmetry moreover provides a solid baseline against which we can compare and contrast results from non-spherical scenarios. \textsc{bamps} is a 3d code, but we make use of the Killing isometries of the spacetime, and employ the Cartoon method~\cite{Alcubierre:1999ab,Pretorius:2004jg}, to suppress two spatial dimensions of the problem. As mentioned above, this allows us to run our spherical jobs on a desktop machine.

\begin{table*}[t!]
   \label{tbl:sph_fam_realim}
    \begin{ruledtabular}
    \begin{tabular}{llllllll}
    \textbf{Family} & \textbf{Gauge} & \textbf{$a^\textrm{Re}$} & \textbf{$r_0^\textrm{Re}$ }& \textbf{$s^\textrm{Re}$}& \textbf{$a^\textrm{Im}$}  &  \textbf{$r_0^\textrm{Im}$}  & \textbf{$s^\textrm{Im}$}\\  \hline
     \textbf{Real}& \textbf{HDWG-$\ln(\alpha)$}&  tuned & 5.0 &1.0  &  0.0 & - & - \\
      \textbf{I: ing}& \textbf{HDWG-$\ln(\alpha)$}&  tuned & 4.0 &2.0 & 0.03 & 3.0 &1.0 \\ 
      \textbf{II: ts}& \textbf{DSS}& 0.01 & 5.0 &1.0 & tuned & 0.0 &2.0 \\  
      \textbf{III: ts}& \textbf{DSS}& -0.1 & 0.0 &2.0 &tuned & 0.0 &1.0 \\ 
    \end{tabular}
    \end{ruledtabular}
      \caption{Initial data for the real and imaginary parts of the complex scalar field of the spherical families. The fact that each family has a different initial configuration but yet gives rise to identical spacetimes at the threshold of collapse serves as a demonstration of universality in spherical symmetry.}
\end{table*}

We consider two types of initial data families for the scalar field, time symmetric data and ingoing data.
The initial data for our time symmetric families have the following functional form
\begin{align}
  \psi^{\textrm{Re}}&=  a^{\textrm{Re}} \exp\left( -\left(\sqrt{\left(\frac{z}{s_{z}^{\textrm{Re}}}\right)^2 + \left(\frac{\rho}{s_{\rho}^ \textrm{Re}}\right)^2} - r_{0}^{\textrm{Re}}\right)^2\right) \label{eq:17} \\
  \notag &+ a^{\textrm{Re}} \exp\left( -\left(\sqrt{\left(\frac{z}{s_{z}^{\textrm{Re}}}\right)^2 + \left(\frac{\rho}{s_{\rho}^ \textrm{Re}}\right)^2} + r_{0}^{\textrm{Re}}\right)^2\right),
\end{align}
for the real part, and, 
\begin{align}
   \psi^{\textrm{Im}}&= a^\textrm{Im} \exp\left( -\left(\sqrt{\left(\frac{z}{s_{z}^{\textrm{Im}}}\right)^2 + \left(\frac{\rho}{s_{\rho}^\textrm{Im}}\right)^2} - r_{0}^{\textrm{Im}}\right)^2\right) \label{eq:18} \\ 
   \notag &+ a^\textrm{Im} \exp\left( -\left(\sqrt{\left(\frac{z}{s_{z}^{\textrm{Im}}}\right)^2 + \left(\frac{\rho}{s_{\rho}^\textrm{Im}}\right)^2} + r_{0}^{\textrm{Im}}\right)^2\right),
\end{align}
for the imaginary part, whilst $\Pi^{\textrm{Re}}=\Pi^{\textrm{Im}}=0$. The Hamiltonian and momentum constraints are solved as discussed in Section~\ref{subsubsection:ID}, with remaining violation of order~$O(10^{-7})$  or better in the initial data once evaluated on the evolution domain.

In the case of ingoing initial data, we, additionally, turn on~$\Pi^{\textrm{Re}}$ and~$\Pi^{\textrm{Im}}$, in such a way that the pulses are mostly incoming. This is achieved by considering spherically symmetric solutions to the flat-space wave equation, imposing relative smallness of~$(\partial_t-\partial_r)(r\psi^\textrm{Re/Im})$ at the start, together with regularity of the data.
The explicit formulas that we use for the time derivatives are the following
\begin{align}
  \Pi^{\textrm{Re}}&=  -2 a^{\textrm{Re}} \left(\sqrt{\left(\frac{z}{s_{z}^{\textrm{Re}}}\right)^2 + \left(\frac{\rho}{s_{\rho}^ \textrm{Re}}\right)^2} - r_{0}^{\textrm{Re}}\right) \notag \\
  &\exp\left( -\left(\sqrt{\left(\frac{z}{s_{z}^{\textrm{Re}}}\right)^2 + \left(\frac{\rho}{s_{\rho}^ \textrm{Re}}\right)^2} - r_{0}^{\textrm{Re}}\right)^2\right)  \notag\\
  \notag &+ 2 a^{\textrm{Re}} \left(\sqrt{\left(\frac{z}{s_{z}^{\textrm{Re}}}\right)^2 + \left(\frac{\rho}{s_{\rho}^ \textrm{Re}}\right)^2} + r_{0}^{\textrm{Re}}\right) \\
  &\exp\left( -\left(\sqrt{\left(\frac{z}{s_{z}^{\textrm{Re}}}\right)^2 + \left(\frac{\rho}{s_{\rho}^ \textrm{Re}}\right)^2} + r_{0}^{\textrm{Re}}\right)^2\right) \label{eq:19},
\end{align}
for the real part, and, 
\begin{align}
  \Pi^{\textrm{Im}}&=  -2 a^{\textrm{Im}} \left(\sqrt{\left(\frac{z}{s_{z}^{\textrm{Im}}}\right)^2 + \left(\frac{\rho}{s_{\rho}^ \textrm{Im}}\right)^2} - r_{0}^{\textrm{Im}}\right) \notag \\
  &\exp\left( -\left(\sqrt{\left(\frac{z}{s_{z}^{\textrm{Im}}}\right)^2 + \left(\frac{\rho}{s_{\rho}^ \textrm{Im}}\right)^2} - r_{0}^{\textrm{Im}}\right)^2\right)  \notag\\
  \notag &+ 2 a^{\textrm{Im}} \left(\sqrt{\left(\frac{z}{s_{z}^{\textrm{Im}}}\right)^2 + \left(\frac{\rho}{s_{\rho}^ \textrm{Im}}\right)^2} + r_{0}^{\textrm{Im}}\right) \\
  &\exp\left( -\left(\sqrt{\left(\frac{z}{s_{z}^{\textrm{Im}}}\right)^2 + \left(\frac{\rho}{s_{\rho}^ \textrm{Im}}\right)^2} + r_{0}^{\textrm{Im}}\right)^2\right) \label{eq:20},
\end{align}
for the imaginary part.

The initial conditions (amplitude~$a^\textrm{Re/Im}$, location of center~$r_0^\textrm{Re/Im}$ and width~$s_{z}^\textrm{Re/Im}=s_{\rho}^\textrm{Re/Im}\equiv s^\textrm{Re/Im}$) for the Gaussian pulses of our spherical families are shown in Table~\ref{tbl:sph_fam_realim}. Tuning either the real or the imaginary part is equivalent, since they take part in the equations of motion identically. To demonstrate this, we are alternating between tuning the real or imaginary part in different families.
We have evolved all families using two different gauges, first that appearing in~\cite{deppe2019critical} and also that of~\cite{cors2023formulation}, which we refer to as HDWG-$\ln(\alpha)$ and DSS-compatible respectively. We only present results from the gauge choice that leads us closer to the threshold of collapse for each family. In general terms the best gauge choice is one that allows to avoid coordinate singularities and lets us tune further. However, it is worth mentioning that the differences between the cleanest results from each gauge are minor. One of the families is ingoing, denoted as ``ing'', whilst the rest start at a moment of time symmetry, denoted as ``ts''. The last half-echo of every family is sensitive to numerical error. The more it resembles the earlier peaks, the better the results are. In the case of the ingoing family we see a slightly better last half-echo when using the HDWG-$\ln\left(\alpha\right)$ gauge rather than the DSS-log gauge and this gauge furthermore allowed us to tune two more digits. For these reasons we chose to present this gauge for that particular family.

\begin{figure}[t!]
    \centering
    \includegraphics[width=\columnwidth]{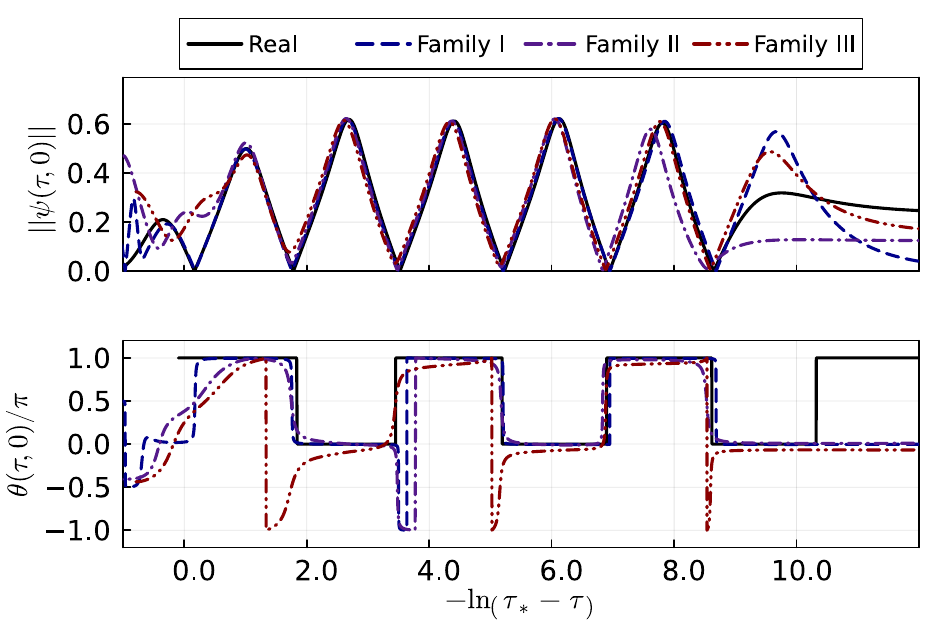}
    \caption{The absolute value of the real scalar field at the center with respect to retarded proper adapted time for different initial conditions, where $\tau$ is the proper time at the origin and $\tau_*$ is the accumulation time of DSS. The upper panel shows the universal behavior of the amplitude of the complex scalar field at late times of the evolution. The lower panel indicates that the phase of the complex field seems to maintain a universal period in spherical symmetry, but not a universal function.}
    \label{fig:Realvscomplex}
\end{figure}

In spherical symmetry critical behavior has been validated for a wide variety of models. Although the picture changes quantitatively for different matter models, many qualitative properties hold in common near the threshold of collapse. In our case, there are two competing scalar fields, namely the real and imaginary parts of the complex scalar field, which only interact through their minimal coupling to gravity. Thus in our case the interesting point to understand is how the solution space of the real field is embedded within that of the complex field. Tuning one real scalar field family to the threshold and comparing that to our three complex field families, the first striking result is that individually, the real and imaginary parts of the field do not appear to follow a universal evolution near the threshold, but instead collectively contribute to the universal critical behavior at late times that is manifest in quantities invariant under rotation through a constant phase. For instance, as demonstrated in  Figure~\ref{fig:Realvscomplex}, the absolute value of the massless complex field does follow a universal evolution near the threshold.

\begin{figure*}[!t]
    \centering    
    \vspace{0cm}
    \includegraphics[scale=0.6]{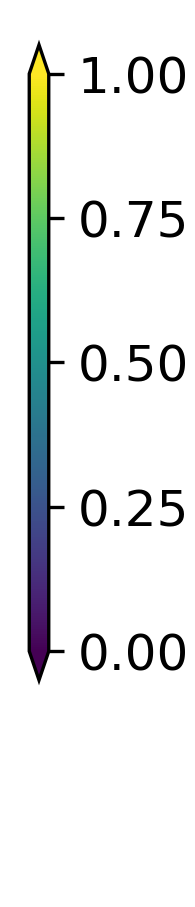} \,
    \includegraphics[scale=0.365]{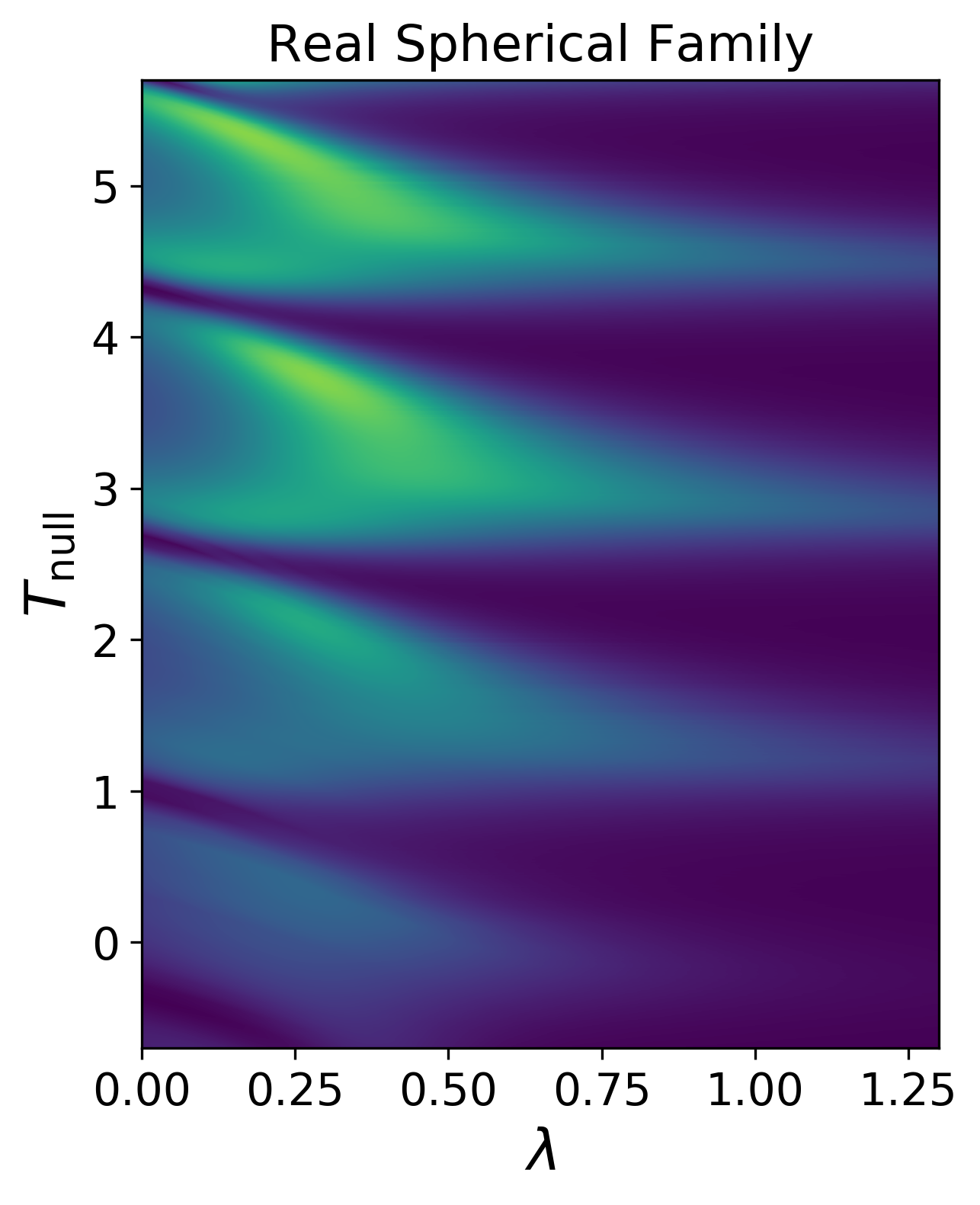} \,
    \includegraphics[scale=0.365]{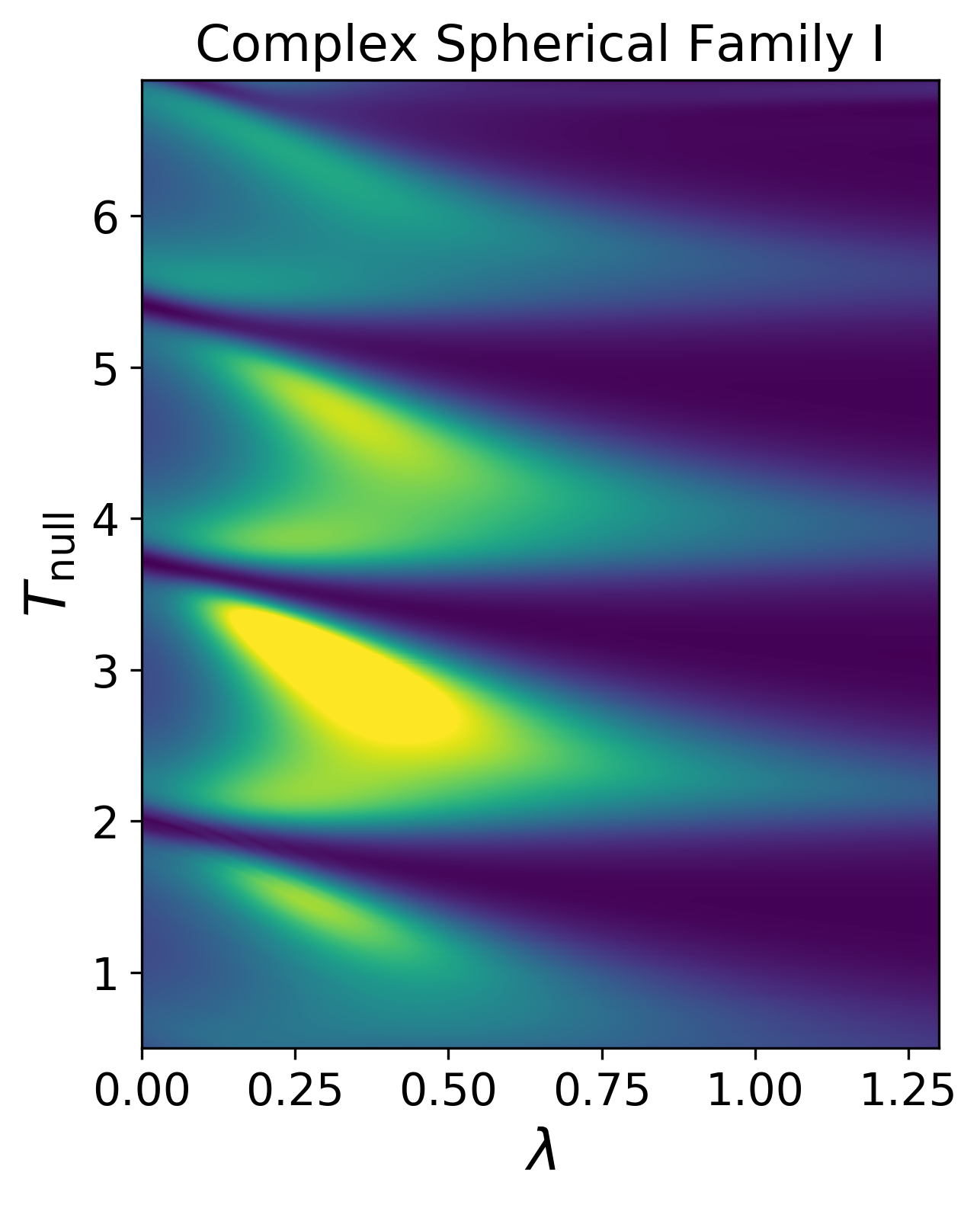} \,
    \includegraphics[scale=0.365]{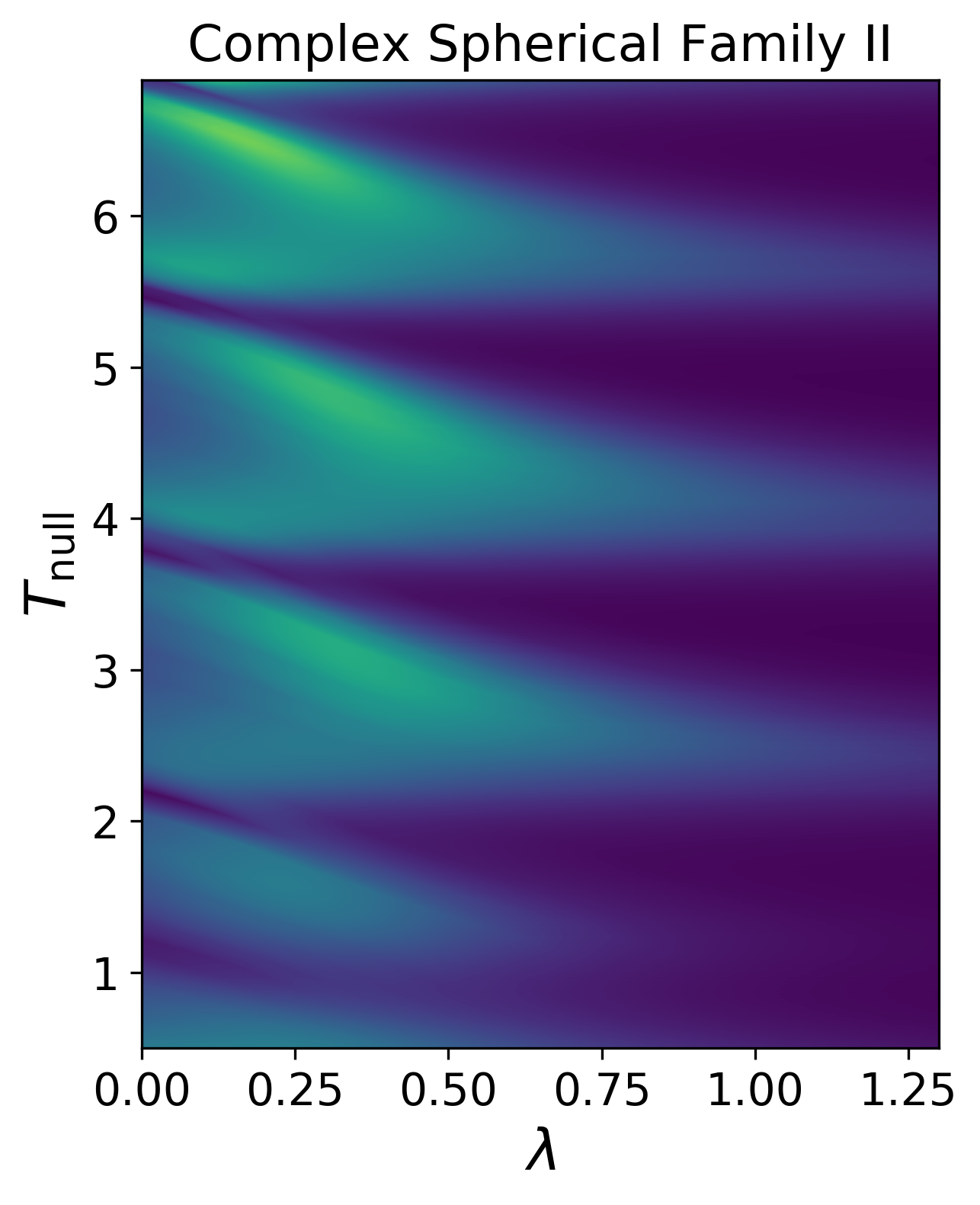} \,
    \includegraphics[scale=0.365]{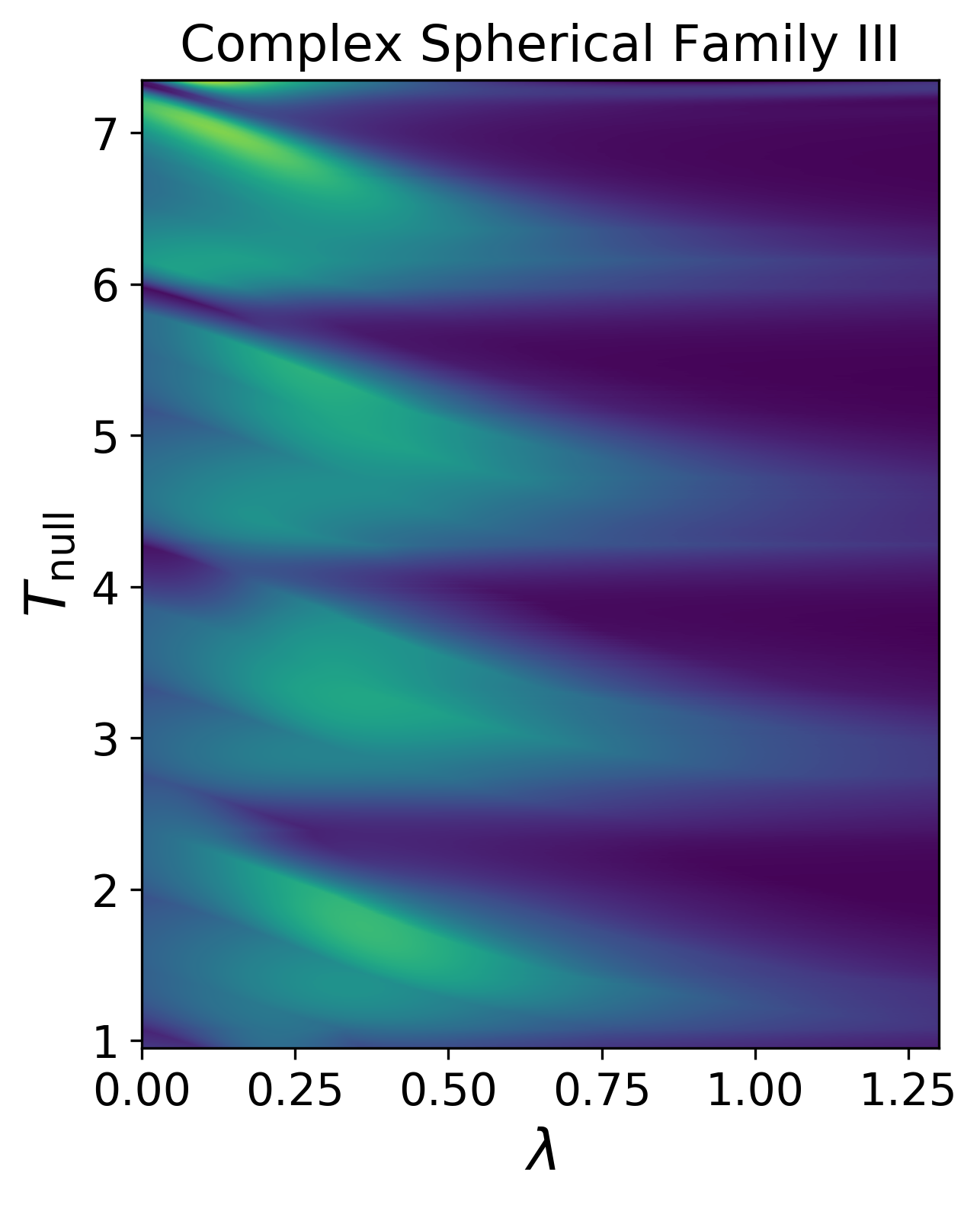} \,
    \caption{Color maps of the rescaled energy density~$(\tau_*-\tau)^2\rho$ as a function of the single-null similarity coordinates~$(T_{\textrm{null}},\lambda)$, see main text for details, for our best-tuned data from each of the spherical families enumerated in Table~\ref{tbl:sph_fam_realim}. The apparent periodicity of each plot in~$T_{\textrm{null}}$ is evidence of the DSS nature of the threshold solutions. The plotted quantity is foliation dependent through the appearance of~$n^a$, but one might expect that similar spacetimes are sliced by the same dynamical gauge condition in a similar manner. In any case, the close similarity of the four plots is evidence that there is a universal critical solution, and that the metric of that spacetime agrees identically with that of the Choptuik solution obtained with a real scalar field.}
\label{fig:rho_nullplots}
\end{figure*} 

To render Figure~\ref{fig:Realvscomplex}, we postprocess the data of the best tuned subcritical evolution from each family to construct a similarity time coordinate. To achieve this we define~$\tau$ as the proper time elapsed at the origin from the initial data,
\begin{align}
    \tau (t) =\int_{0}^{t} \alpha(t', 0)\,  dt' \, ,
    \label{eq:proper_time}
\end{align}
where~$\alpha$ here is the lapse function, and calculate the accumulation time, $\tau_*$, of the DSS behavior making use of two pairs of zero crossings of the real part of the complex scalar field~$\left(\tau_n, \tau_{n+1}\right)$ and~$\left(\tau_m, \tau_{m+1}\right)$.
We calculate $\tau_*$ through an estimation formula as it appears in~\cite{baumgarte2018aspherical},
\begin{align}
    \tau_* = \frac{\tau_n\tau_{m+1}-\tau_{n+1}\tau_{m}}{\tau_n- \tau_{n+1}-\tau_m +\tau_{m+1}}\,.
    \label{eq:taustar}
\end{align}
The echoing period, $\Delta$, if calculated using the real or imaginary parts separately is the following
\begin{align}
    \Delta=2\ln{\frac{\tau_*-\tau_n}{\tau_*-\tau_{n+1}}}\,,
    \label{eq:delta}
\end{align}
due to the fact that for each pair of zero crossings half a period~$\Delta/2$ has elapsed. To extract the period from the complex families of Table~\ref{tbl:sph_fam_realim}, we employ the absolute value of the complex scalar field and subtract the mean value of the DSS-periodic part for these spherical runs to allow the curve to have zero crossings. Obviously, this curve has half the period of a real scalar field. Thus, if we use the zero crossings of such a curve in~\eqref{eq:delta}, we will need a pair of zero crossings of the absolute value of the following form~$\left(\tilde{\tau}_n, \tilde{\tau}_{n+2}\right)$ in order to retrieve a period comparable to the one of the Choptuik solution.
\begin{align}
    \Delta=2\ln{\frac{\tau_*-\tilde{\tau}_n}{\tau_*-\tilde{\tau}_{n+2}}}\,.
    \label{eq:delta2}
\end{align}
Averaging over all $n$ within each spherical family and over all complex families the value we find is~$\bar{\Delta}=3.43\pm 0.03$. The relevant value for our real scalar field run, using data appearing in~\cite{cors2023formulation}, is $\Delta = 3.43\pm 0.03$. Since we plot the absolute value of the complex field, superficially the upper panel of Figure~\ref{fig:Realvscomplex} appears to show more repeated features than in the standard plots for the real scalar field. But six peaks of the absolute value correspond to three DSS-echoes in a standard plot. The natural point of comparison in the literature for Figure~\ref{fig:Realvscomplex} is instead Figure~$2$ of~\cite{jimenez2022critical}. Although~\cite{jimenez2022critical} studied the massive field, the data plotted in their Figure~$2$ were in the regime where the effect of the mass was expected to be negligible (since length scales are smaller than Compton wavelength). We therefore take the excellent agreement of our plot with theirs as evidence that our data are correct. In the lower panel of Fig.~\ref{fig:Realvscomplex} we have plotted the phase of the complex scalar field for each family, which seems to alternate, roughly, between two constant values during the evolution and presents repeated features with a universal period. In order to examine whether there is a universal phase we have used exactly the same alignment in similarity time in the upper panel of the same figure, but we have added an overall constant phase, $\vartheta_0\text{\,mod\,} 2 - 1$,  to each of the curves. According to this plot, the phase exhibits periodic features with a universal period, yet there is no evidence that it follows a universal solution. We have additionally examined various quantities, including for instance the Ricci scalar, that are invariant under constant changes to the phase of the complex field at the origin as a function of slow-time, and find in all cases that they agree well with the values of the Choptuik solution. 
Looking instead at the phase of the complex field at the origin, $\theta=\arctan\left(\psi^{\textrm{Im}}/\psi^{\textrm{Re}}\right)$, as a function of slow-time there is no obvious universal structure across the four families (see Figure~\ref{fig:Realvscomplex}).
For a pure real or imaginary configuration the phase would, of course, remain constant throughout the evolution. By the~$U(1)$ symmetry of the field equations, there are therefore solutions with arbitrary constant phase. For our generic configurations however, near the threshold, the phase is approximately periodic within each family, remaining approximately constant through segments of each scale echo, before quickly shifting to another constant value.

For subcritical runs we also find the scaling exponent~$\gamma$ of the power-law,
\begin{align}
\label{eq:Rmaxequation}
    R_{\text{max}}\sim \left(a-a_*\right)^{-2\gamma}\,,
\end{align}
fitting the values by regression. We find values around~$\gamma\simeq 0.37$ for all families. This coincides with the familiar value from simulations of the real scalar field in spherical symmetry~\cite{choptuik1993universality}. 

Based on the evidence presented above, it is tempting to conjecture that the metric of the spherical critical spacetime for the complex scalar field agrees exactly with that of the Choptuik solution. Yet more evidence is given for this finding by constructing single-null DSS-adapted coordinates, following the steps described in~\cite{Baumgarte:2023saw,baumgarte2023critical}. In particular we extend the slow-time used for Figure~\ref{fig:Realvscomplex}, which we now call~$T_{\textrm{null}}$, as
\begin{align}
    \label{eq:slow_time_null}
    T_{\textrm{null}}= -\ln(\tau_*-\tau) \,,
\end{align}
where again~$\tau$ is computed from Eq.~\eqref{eq:proper_time} and~$\tau_*$ from Eq.~\eqref{eq:taustar}. This is done by integrating out null geodesics from the origin in the radial direction, using an affine parameter~$\lambda$ normalized so that~$d\lambda/d\tau=(\tau_*-\tau)^{-1}$ at the center, and labeling the geodesics by their value of~$T_{\textrm{null}}$ at the center. These null curves are computed in post-processing on a grid uniform in code coordinates. Due to the interpolation and integration error on that grid combined with the sensitive dependence of~$T$ through the logarithmic term in its definition, we do find it necessary to shift our values of~$\tau_*$ slightly. In case the spacetime is DSS with accumulation point at the center, the coordinates~$(T_{\textrm{null}},\lambda)$ will be adapted to the symmetry. Therefore, plots of dimensionless scalars in these coordinates, such as suitably rescaled curvature scalars or components of the stress-energy tensor, should be periodic in~$T_{\textrm{null}}$. We have rendered such heat maps and, upon doing so, we see strong evidence of this periodicity. In Fig.~\ref{fig:rho_nullplots} for instance, we plot the rescaled energy density~$(\tau_*-\tau)^2\rho$ as a function of similarity coordinates for the best tuned spacetime obtained within each of our four spherical families of initial data. (Recall that~$\rho=n^a n^bT_{ab}$). We also observe remarkable agreement between the threshold solutions of all four families in all such quantities. Again, an example is given by the four panels of Fig.~\ref{fig:rho_nullplots}. As these plots lie over an extended spatial region, they demonstrate much more strongly the agreement of the threshold solutions of each family than Fig.~\ref{fig:Realvscomplex}.

\subsection{Departures from Spherical Symmetry}

\begin{table*}[!t]
\label{tbl:axi_families}
    \begin{ruledtabular}
    \begin{tabular}{ccccccccccc}
    \textbf{Family}  & $\epsilon^2$& \textbf{$s_{z}^{\textrm{Re}}$} & \textbf{$s_{\rho}^{\textrm{Re}}$}& $a^{\textrm{Re}}_*$ & $\tau_*$ & $\gamma$ & $\Delta_1$& $\Delta_2$ & $\Delta_3$ &$\bar{\Delta}$\\ \hline
      \textbf{I}& $0$ & $1.0$ & $1.0$& $0.1140412$ &$1.570$ & $0.370\pm0.002$  & $3.45\pm 0.03$&$3.44\pm 0.02$&$3.42\pm 0.02$&$3.44\pm 0.02$\\
      \textbf{II}& $0.25$ &$1.1547$ & $1.0$& $0.1142897$ &$1.614$ & $0.372\pm0.002$ & $3.45\pm 0.02$& $3.47\pm0.03$&$3.37\pm0.06$&$3.43\pm 0.02$\\
      \textbf{III}& $ 0.75$ &$2.0$& $1.0$& $0.1187356$ & $1.811$& $0.364\pm0.002 $ & $3.38\pm0.02$&$3.39\pm0.04$&$3.323\pm0.014$&$3.36\pm0.02$\\
      \textbf{IV}& $0.99$ &$10.0$ & $1.0$& $0.1328560$ & $2.667$ & $0.329\pm0.002 $ & $2.95\pm 0.14$&$3.09\pm 0.09$&$2.97\pm0.48$&$3.0\pm0.2$\\
    \end{tabular}
     \end{ruledtabular}
    \caption{Initial data and critical parameters as calculated from the most tuned simulations of all the aspherical families ($\epsilon^2>0$), in comparison to the spherical one ($\epsilon^2=0$). In fact, we alter $s_z^\textrm{Re}$ in order to parametrize the asphericity. The use of $\epsilon^2$ here is made for naming purposes, so that it agrees with \cite{baumgarte2018aspherical}, and the relation between $s_z^\textrm{Re}$ and $\epsilon^2$ is given by \eqref{eq:id_baumgarte_relation}. All initial data families correspond to a moment of time symmetry and are evolved with the DSS-compatible gauge. We tune~$a^\textrm{Re}$ towards the threshold of collapse, $a^\textrm{Re}_*$, with~$11$ digits accuracy, whilst we set $r_0^\textrm{Re}=0.0$. The initial data for the imaginary part is simply a spherical gaussian configuration with~$a^\textrm{Im}=0.1$, $r_0^\textrm{Im}=0.0$, $s_{z}^\textrm{Im}=s_{\rho}^\textrm{Im}=1.0$. The axisymmetric contribution comes from the real part of the complex scalar field initially, and gets inherited by the imaginary part during the evolution, since they interact indirectly via their minimal coupling to gravity. We calculate the DSS-echoing period in three different ways, using the periodicity of the central values of the scalar field~$\Delta_1$, the global maxima of the Ricci scalar against the distance in phase space from the critical solution~$\Delta_2$, and of the maxima of the Ricci scalar curvature plotted against similarity time~$\Delta_3$. Observe that the error bounds quoted here stem solely from our regression, and should not be taken as a reliable indication of the accuracy of the underlying numerical spacetime data.
    }
\end{table*}

Treating now spacetimes that are only axisymmetric, we can use the Cartoon method to suppress just one spatial dimension. We follow a similar approach to~\cite{choptuik2003critical,baumgarte2018aspherical}, who report a gradual drift in the power-law and echoing parameters as the degree of asphericity increases, followed by a bifurcation of the center of collapse away from the origin with large enough~$\epsilon$. The reported bifurcation occurs both parametrically in solution space and in time for single spacetimes. The centers of collapse appear as distinct peaks of curvature that could correspond to a single or two DSS-accumulation points, or to neither. Further investigation is needed to establish whether or not each individual center locally approximates the spherical critical solution. On the other hand, in the only axisymmetric work on massless complex scalar field critical collapse~\cite{choptuik2004critical}, which included non-vanishing angular momentum, these features do not manifest. Presently, we focus on the former phenomenology, but in the future it will be important to consider also the transition to include angular momentum also.

The initial data for our aspherical families are obtained by Eqs.~(\ref{eq:17}) and (\ref{eq:18}) by setting $r_0^\textrm{Re/Im}=0$ and values of $s_z^\textrm{Re/Im}$ and $s_\rho^\textrm{Re/Im}$ as shown in Table~\ref{tbl:axi_families}. The translation of these initial data to the notation appearing in~\cite{baumgarte2018aspherical} is straightforward. In particular, we have,
\begin{align}
\label{eq:id_baumgarte_relation}
    s_z^\textrm{Re}= \sqrt{\frac{1}{1-\epsilon^2}}\, , ~~~~ \, a^\textrm{Re}=\frac{\eta}{2} \,,
\end{align}
with the notation of~\cite{baumgarte2018aspherical} appearing on the right-hand side of the expressions. Even though we are controlling the deviations from sphericity by altering~$s_z^\textrm{Re}$ in the initial data, we will be labeling our families using the parameter~$\epsilon^2$, introduced in~\cite{baumgarte2018aspherical}, for reasons of convenience of comparison. Again, the Hamiltonian and momentum constraints are solved as discussed in Section~\ref{subsubsection:ID}, with remaining violation of order at worst~$O(10^{-6})$ in the initial data once evaluated on the evolution grid.

\begin{figure}[!t]
    \centering
    \includegraphics[width=\columnwidth]{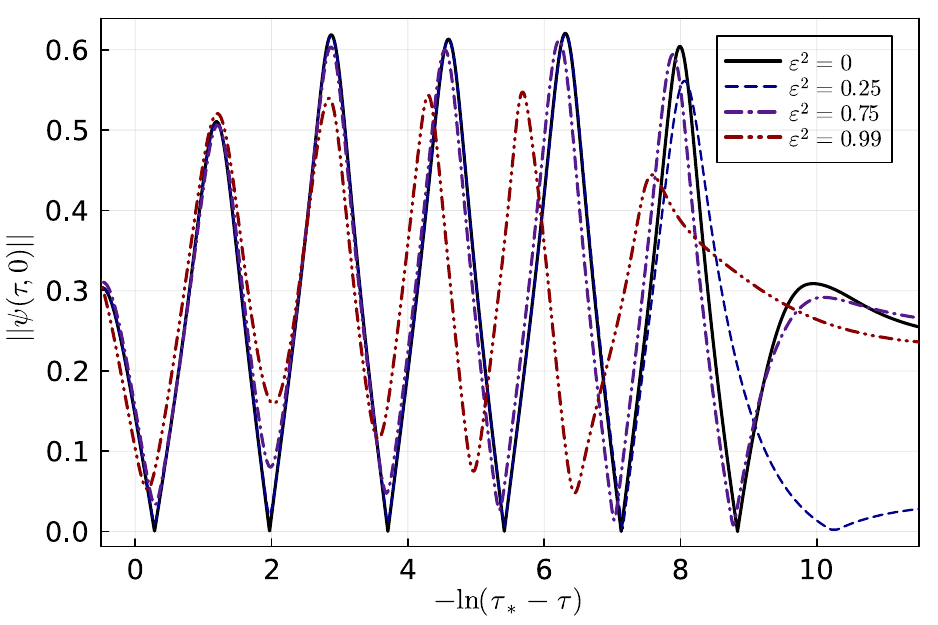}
    \caption{The central absolute value of the complex scalar field in similarity coordinates for the most tuned evolutions of all aspherical families. All families have been comparably tuned and have been aligned horizontally in such a way that the first period for each one of the curves lies on top of each other. This brings confidence that the drift in the echoing period that we can readily see by eye inspection in this plot is physical rather than a numerical artifact.}
    \label{fig:axivssph}
\end{figure}

\begin{figure*}[!t]
    \centering
    \includegraphics[width=\columnwidth]{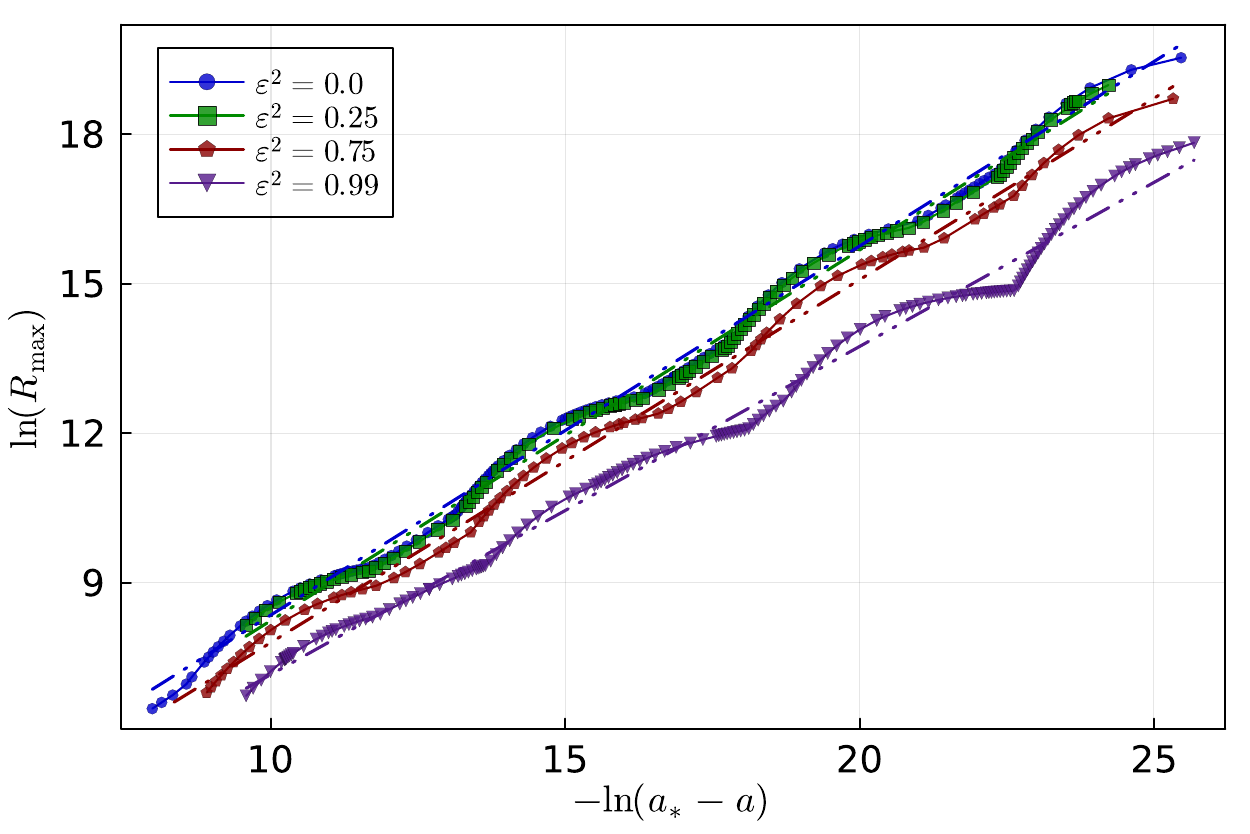} \, \,
    \includegraphics[width=\columnwidth]{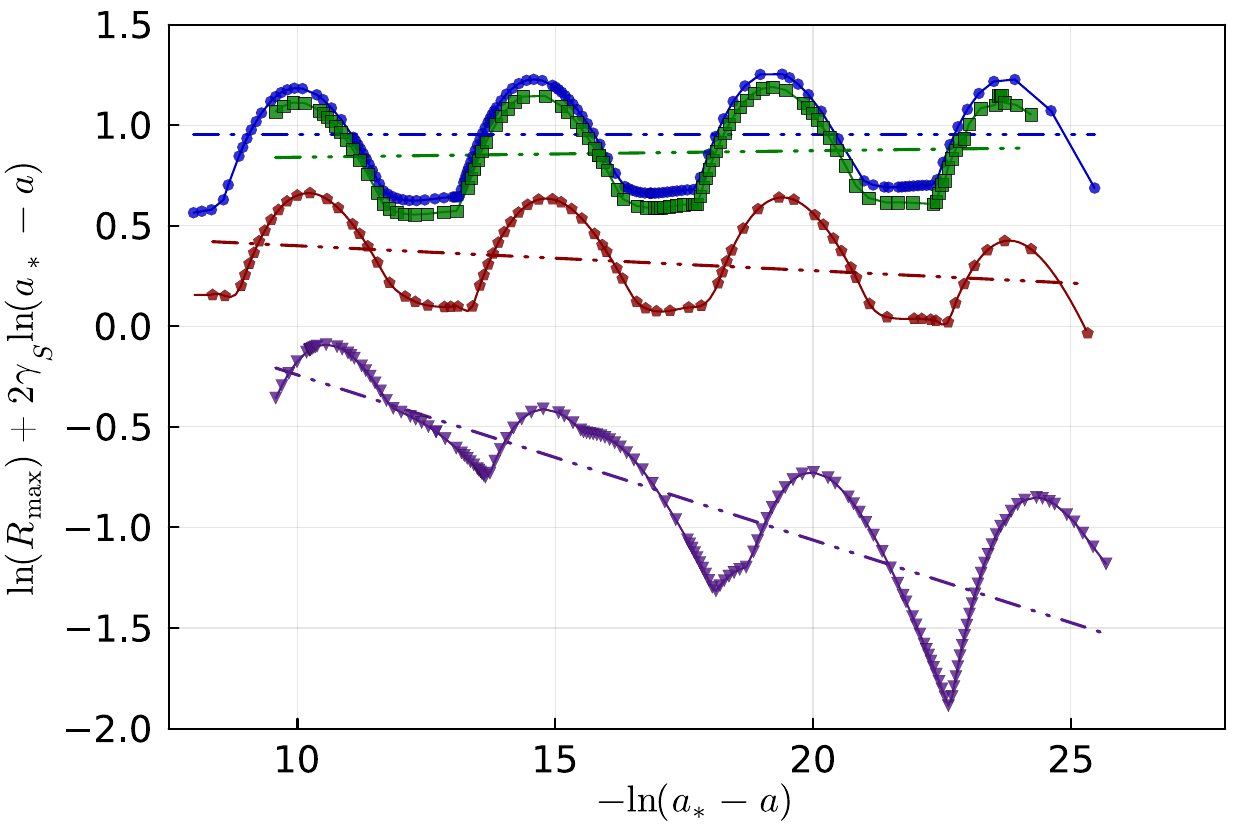}
    \caption{On the left we plot the natural logarithm of the global maximum in space and time of the Ricci scalar curvature invariant against the logarithm of the difference of the amplitude with respect to our estimation for the critical amplitude. One can observe a similar power-law behavior for all curves and a clear periodicity of the DSS structure for the ones that are close to being spherically symmetric. For the ones that deviate a lot from spherical symmetry we cannot safely claim that there is a clear period nor that the period is universal. On the right plot we subtract the linear contribution of the spherical simulation (see Eq.~\ref{eq:Rmaxequation} where we call the exponent for the spherical solution~$\gamma_{\text{S}}$) from all curves of the left panel to demonstrate more evidently the downward trend of the power law scaling exponents with increasing $\epsilon^2$, as shown in \cite{baumgarte2018aspherical} for the real case as well.}
    \label{fig:axiscalingwl}
\end{figure*}

Numerical data produced with~\textsc{bamps} has been thoroughly convergence tested in numerous challenging validation runs, see for instance~\cite{hilditch2016pseudospectral,renkhoff2023adaptive,Atteneder:2023pge}. Doing so carefully is a technically tricky task with any pseudospectral method, because one cannot expect a simple power-law improvement in errors with resolution as in a finite-differencing method. Instead, in the best case scenario one should obtain exponential convergence with resolution, but typically with an unknown exponent. (To avoid misinterpretation: exponential convergence is highly desirable, but complicates quality evaluation). These difficulties are amplified yet further when using mesh-refinement, because as resolution is increased, the grid hierarchies will only match if carefully specified at the outset, which requires more book-keeping to set up. For applications in critical collapse, the situation is yet worse because, by design, we consider spacetimes which depend sensitively on the chosen initial data. Consequently numerical error can have an outsized effect on the outcome of the evolution and can, for instance, even kick the end-state of the evolution from dispersion to black hole formation, and thus end up with quite different levels of constraint violation. For these reasons, and the fact that to do so would be computationally highly expensive, we have not performed {\it systematic} convergence tests on all of the data presented here. In particular, our mesh-refinement infrastructure was tested in detail in~\cite{renkhoff2023adaptive}, including with pre-prescribed grid hierarchy. In~\cite{Atteneder:2023pge} detailed validation runs were presented with exactly the complex scalar field code we employ presently. For this project specifically we performed such tests for strong data far from the threshold of collapse and saw convincing convergence with resolution. Confidence in the quality of our aspherical near-threshold evolutions is bolstered by a number of facts. First, we monitor constraint violation throughout. The constraints do grow as the dynamics become increasingly violent at the threshold of collapse, but they remain at least five orders of magnitude, and often more, smaller than any curvature scalars with equivalent units that we consider. Second, we have already compared above our spherical runs with the (admittedly limited) available literature for the complex scalar field, and find excellent agreement. Below we will reveal that our results agree qualitatively with published work for the analogous setup for the real scalar field. All these points considered, we are therefore confident in the qualitative behavior we find, especially when considering small asphericities, but do not feel justified in stating strict error bounds, which are left for future work. To do so we will need to perform detailed convergence tests, which is complicated because of the sensitive dependence of our evolutions on initial data, the use of AMR, which is essential in this context, and our pseudospectral approximation, for which the exact rate of decay of error with resolution is hard to establish a priori.

\begin{figure*}[!t]
    \centering
    \includegraphics[scale=0.9]{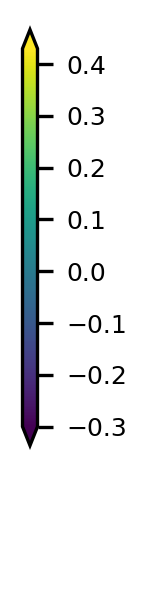}
    \includegraphics[width=0.95\columnwidth]{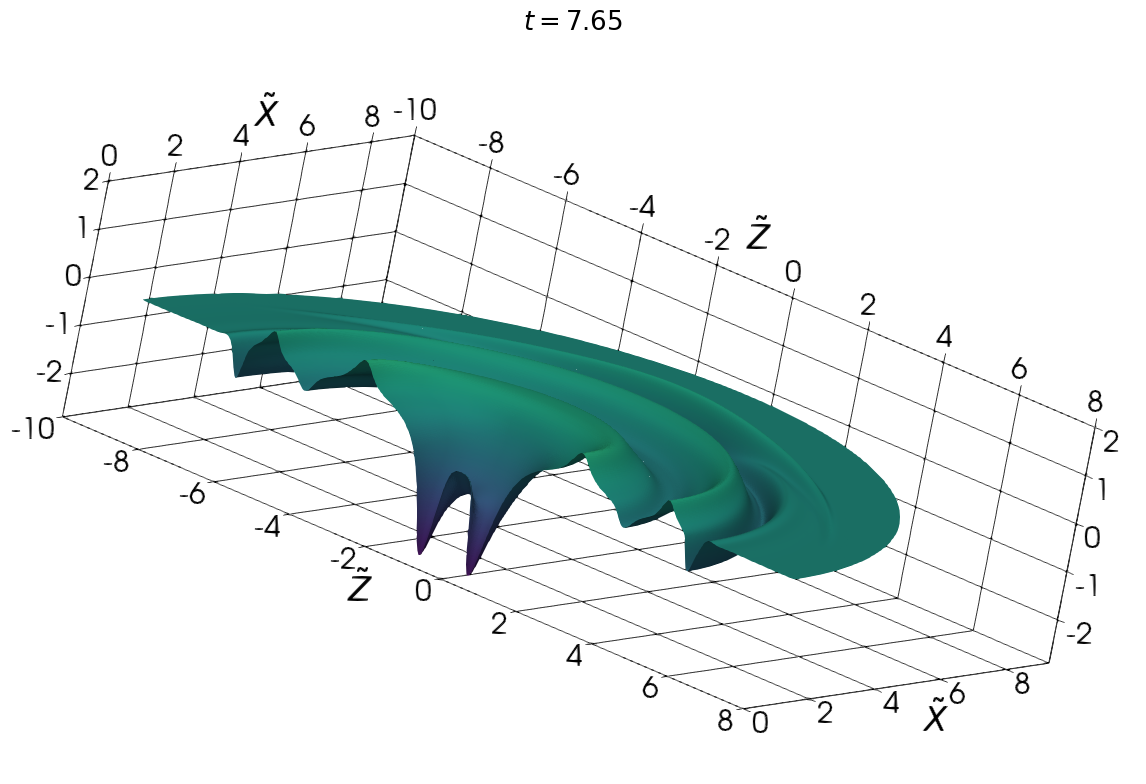}
    \includegraphics[width=0.95\columnwidth]{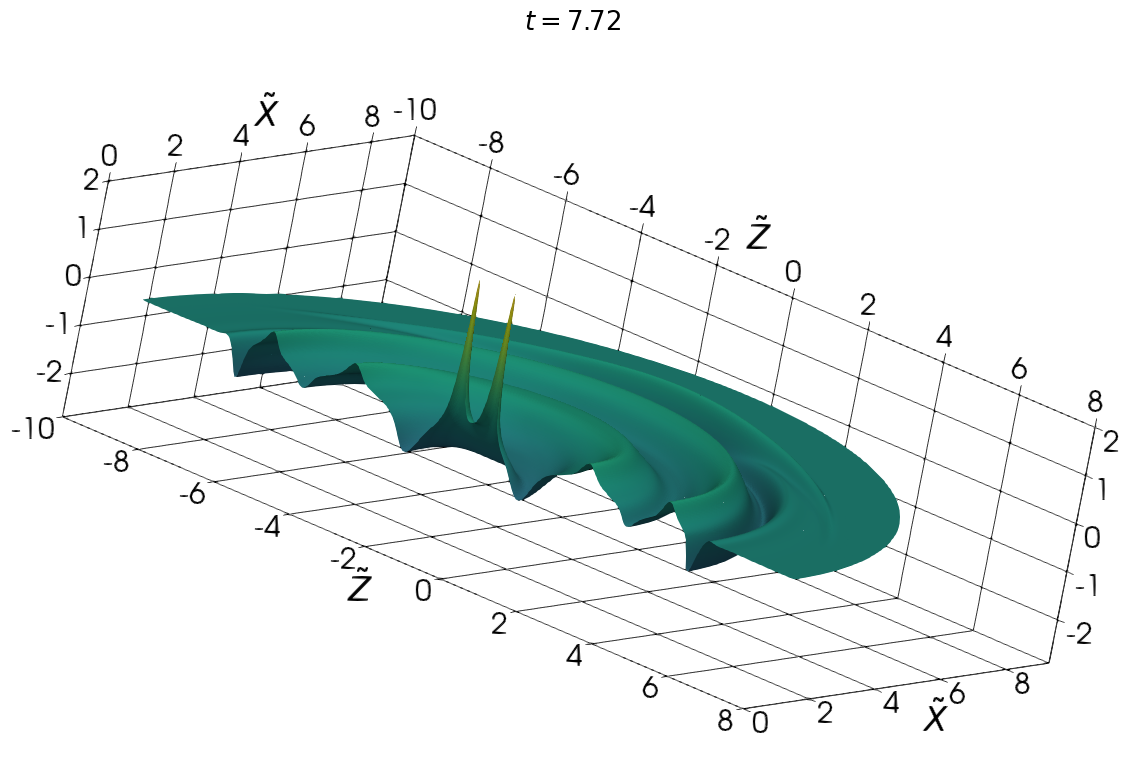}
    \caption{
    Different snapshots of the real part of the complex scalar field at late times of our most tuned and most aspherical evolution $\epsilon^2=0.99$. We have rescaled the axes coordinates to $\tilde{X}=\ln\left(1+100\sqrt{X^2+Z^2}\right)X/{\sqrt{X^2+Z^2}}$ and $\tilde{Z}=\ln\left(1+100\sqrt{X^2+Z^2}\right) Z/{\sqrt{X^2+Z^2}}$, in order to be able to visualise all of the echoes in one figure. The vertical axis shows the rescaled value of the field by a factor of $7$, but the colors do correspond to the actual values of the field. The first snapshot is at $t=7.65$, whilst the second snapshot is at $t=7.73$, where $t$ is the coordinate time of the code. We observe that at late times and for very high tuning, the center of collapse experiences a bifurcation. It is the first time this bifurcation is observed in the complex scalar field case.}
    \label{fig:bifurcation}
\end{figure*}

Moving now to discuss our actual results, Table~\ref{tbl:axi_families} reveals a decrease in the echoing period~$\Delta$ and in the power-law exponent~$\gamma$ with increasing deviation from sphericity, where values were obtained by regression as described in the caption. The value of~$\Delta_1$ has been calculated by virtue of Eq.~\eqref{eq:delta2}, corresponding to the period of the central absolute value of the complex scalar field against similarity adapted time at the origin. The average value of the DSS-periodic part of the time series has been subtracted from the data so that the curve, as it appears in Fig.~\ref{fig:Realvscomplex}, has zero-crossings that can be used to apply the formula of Eq.~\eqref{eq:delta2}. In a similar fashion, for~$\Delta_3$, the curve that represents the global maxima of the Ricci scalar against similarity time at the center has been used (we do not include this plot in the paper). This time, the local minima of that curve have been used instead of zero-crossings in order to determine the period given by Eq.~\eqref{eq:delta2}. Several pairs of consecutive zero crossings or local minima have been used and Table~\ref{tbl:axi_families} contains the resulting average values and errors from these calculations. The value of~$\Delta_2$ represents the period of the Ricci scalar maxima in phase space as depicted in the right panel of Fig.~\ref{fig:axiscalingwl} and has been estimated with a periodic sinusoidal fit. The relevant errors that appear in the table for~$\Delta_2$ are related to the error coming from the non-linear least square fit. Finally, the average value, $\bar{\Delta}=\left(\Delta_1+\Delta_2+\Delta_3\right)/3$, of the three is also present in the same table. Our results are in qualitative agreement with those of~\cite{choptuik2003critical,baumgarte2018aspherical}. A quantitative evaluation of the level of agreement with~\cite{choptuik2003critical,baumgarte2018aspherical} is difficult for two main reasons. As discussed above, we do not have strict error estimates from either our or the earlier numerical computations. Nevertheless, in Table~\ref{tbl:axi_families} we quote errors as estimated purely from the regression. To the extent that they can be taken seriously, the trend is significant. Making a side-by-side comparison of the level of drift between our data and that of~\cite{baumgarte2018aspherical}, it seems that at comparable values of~$\epsilon^2$ there is a smaller drift in the parameters of our data. This may be caused by different levels of numerical error. But since we specify spherically symmetric initial data in the imaginary profiles and then place the aspherical perturbation in the real ones, one would anticipate that at fixed~$\epsilon^2$ our data are closer to spherical symmetry than those of~\cite{baumgarte2018aspherical}. Thus our data are presently consistent with the complex scalar field being subject to the same qualitative and quantitative parametric deviation observed for the real scalar field. This will have to be carefully assessed in the future with a more satisfactory measure of asphericity.

In Figure~\ref{fig:axivssph} we plot the absolute value of the scalar field as a function of slow-time at the center, computed exactly as in Figure~\ref{fig:Realvscomplex}, but now for our aspherical families outlined in Table~\ref{tbl:axi_families}. For a good portion of the plot, the spherical and~$\epsilon^2=0.25$ curves are indistinguishable by eye, but as~$\epsilon^2$ increases beyond that, a monotonic decrease of the period is clearly visible. Interestingly, we see evidence for a reduction in the amplitude of the field with each scale echo besides, and at the largest asphericity the oscillatory features could be described, at best, as approximately periodic as there is considerable drift in the amplitude from one oscillation to the next. In Figure~\ref{fig:axiscalingwl}, rather than looking at the best-tuned spacetime from each family, we consider the behavior of the maximum of the logarithm of the scalar curvature~$\ln(R_\textrm{max})$ as we move parametrically through the solution space, varying~$-\ln(a_*-a)$. In the left we again see that results from the spherical and~$\epsilon^2=0.25$ families are indistinguishable by eye, whereas larger values of~$\epsilon^2$ induce a visible drift in the slope of the curves and the period of the oscillations.

\begin{figure*}[!t]
    \centering    
    \vspace{0cm}
    \includegraphics[scale=0.825]{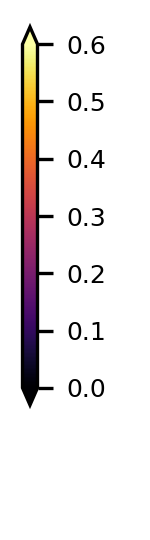}
    \, \includegraphics[width=0.624\columnwidth]{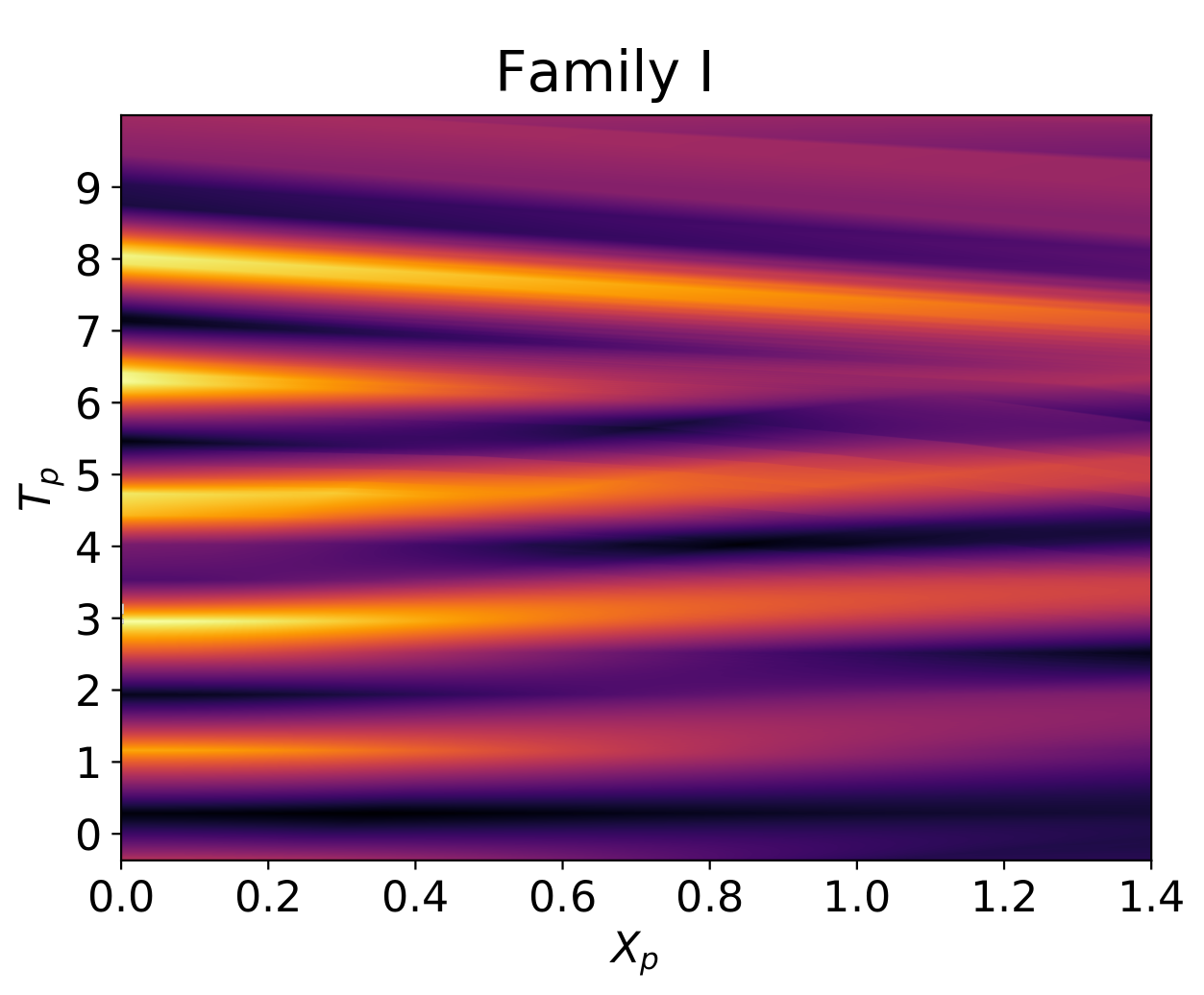} \,
    \includegraphics[width=0.624\columnwidth]{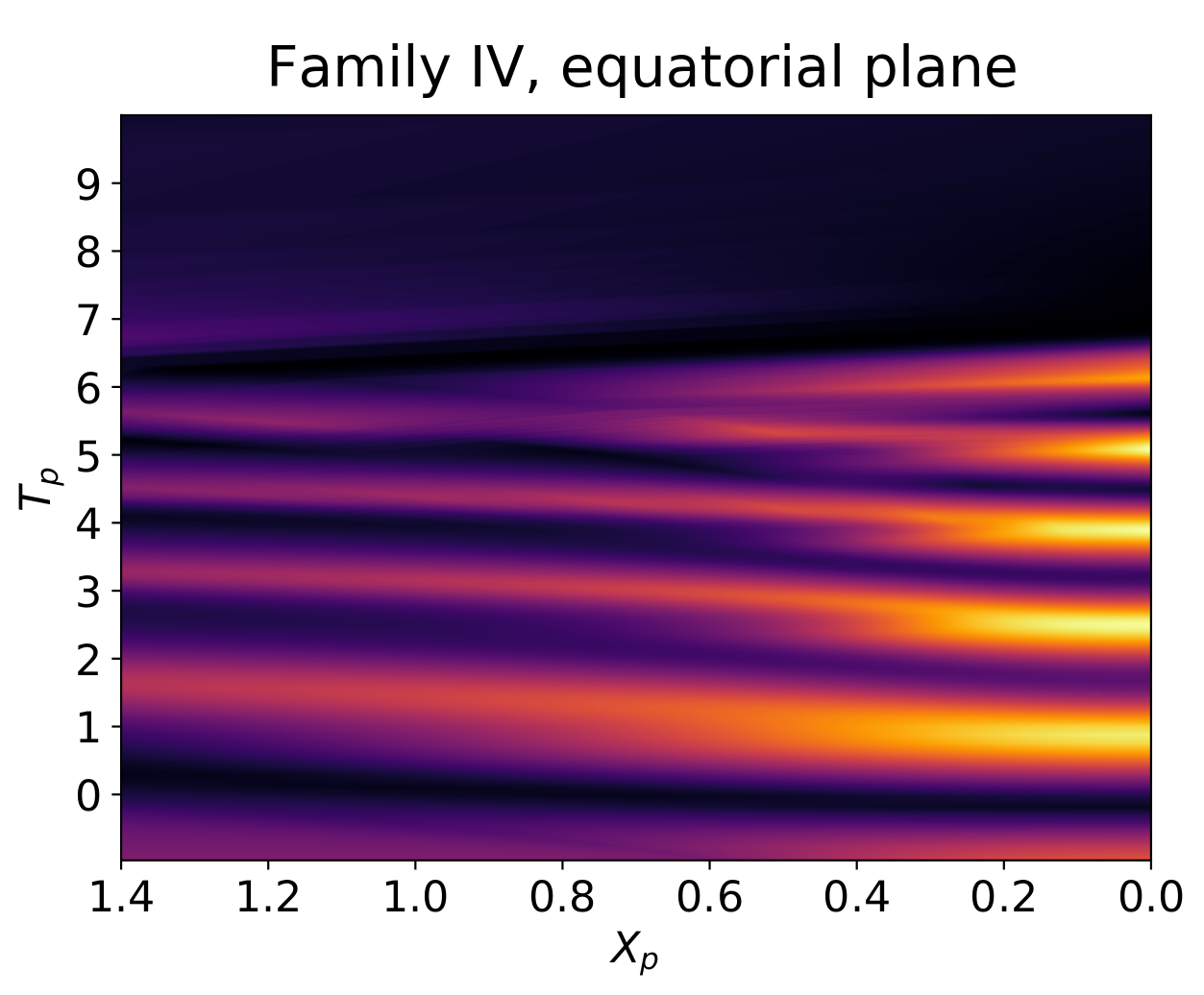}
    \includegraphics[width=0.626\columnwidth]{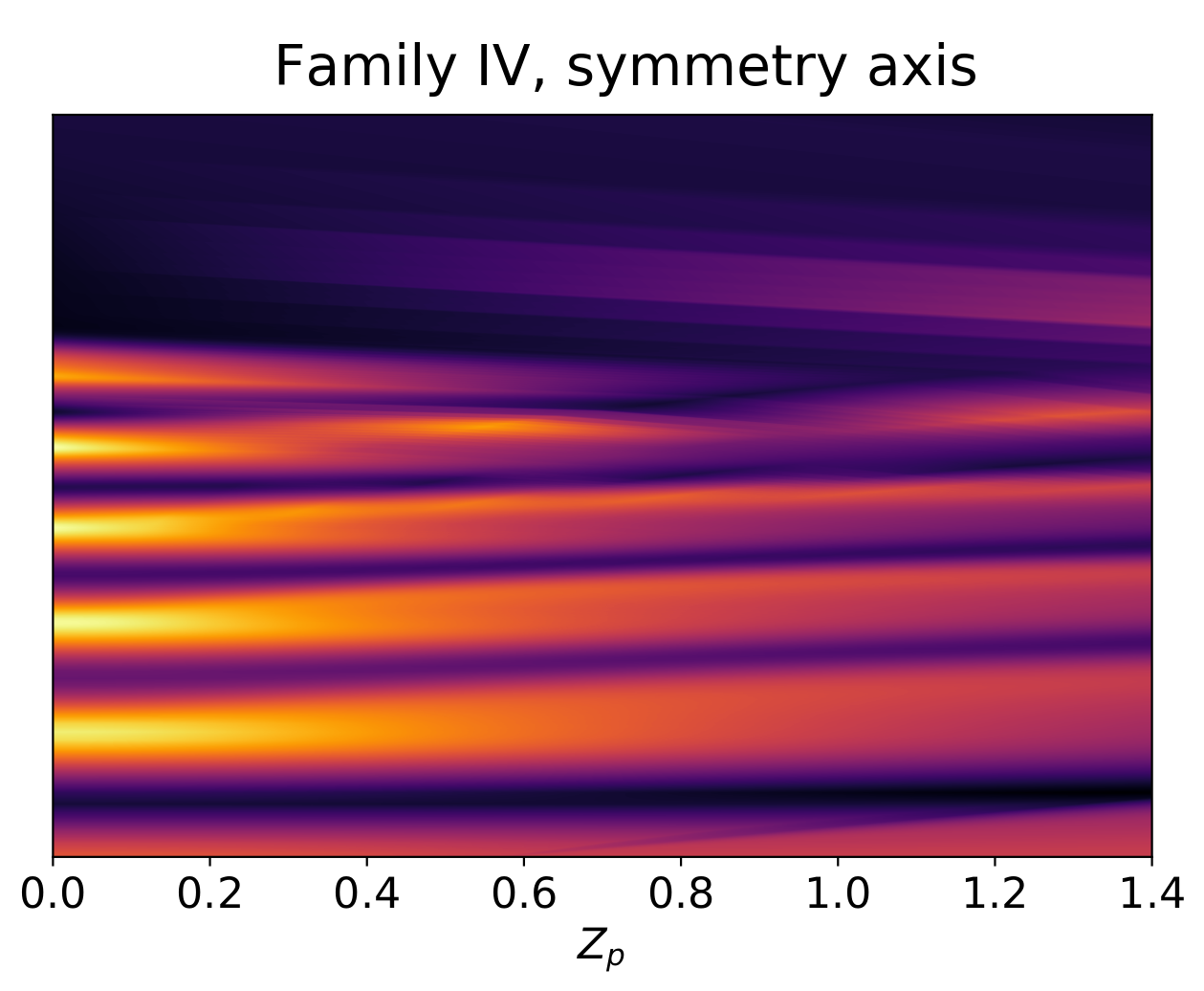}
    \caption{Absolute value of the complex scalar field on the~$z$-axis in coordinates constructed as an attempt to adapt to self-similarity, $Z_p$ is constructed by rescaling the proper distance along the~$z$-axis~\eqref{eq:proper_length} and~$T_p$ is the slow-time coordinate constructed from proper time at the origin, and used here to label slices of the generalized harmonic foliation. On the left we plot data from the spherical run~$\epsilon^2=0$. The lack of periodicity in~$T_p$ is evidence that the foliation is not adapted to DSS. On the right we plot, data from the axisymmetric run with~$\epsilon^2=0.99$. Comparing the two plots one can see bright peaks of the field appearing away from the origin in the~$Z_p$ direction in the axisymmetric run at late times, indicating the bifurcation. A difference in the period of the oscillations at the center is evident.}
    \label{fig:properlength}
\end{figure*}

With very high asphericity, in particular in Family IV which has~$\epsilon^2=0.99$, we see that, with fine-tuning to the threshold beyond around~$7$ digits in~$\alpha$, two centers of collapse form on the symmetry axis, away from the origin. These centers are defined by considering local maxima of non-vanishing curvature scalars and by the dynamical profile of the scalar field itself, as shown in Figure~\ref{fig:bifurcation}, were we plot against renormalized coordinates so that the outgoing pulses can be seen simultaneously. This is the first time that such a bifurcation of the collapse region has been observed for the complex scalar field model, but similar behavior has been observed for the real scalar field and in axisymmetric vacuum collapse.
In particular, both~\cite{choptuik2003critical,baumgarte2018aspherical},  observed such a bifurcation.
The initial data of these references are constructed as a predominantly spherical profile plus a quadrupolar~$Y_{20}$ spherical harmonic perturbation in the scalar field, very similar to ours.
Interestingly, the recent study of~\cite{Reid:2023jmr}, in which instead a dipolar~$Y_{10}$ perturbation was placed, and at a similar level of fine-tuning was arrived at, did not observe either the drift in parameters or the bifurcation. The fully 3d studies of critical collapse~\cite{Healy:2013xia,deppe2019critical} have also not observed the bifurcation, but since the level of fine-tuning achieved in those works is less high, their data are not in contradiction with the bespoke axisymmetric evolutions. Likewise the 3d work of~\cite{martin1999all} does not see evidence for this behavior. Although perplexing, since the study was linear in the perturbation, the latter is not in contradiction with the nonlinear studies.

In a spacetime that contains a curvature singularity, the spatial position at which the maximum curvature is observed is foliation dependent. But since we observe the bifurcation in subcritical data, we can be sure that the basic phenomenon described above will hold regardless of the foliation we work with. This is because if peaks in curvature scalars appear away from the equatorial plane in one foliation, the same must be true in any alternative. All of the data we consider in this work are reflection symmetric about~$z=0$ and thus have a well-defined equatorial plane. Nevertheless, the formation of two centers of collapse could encompass various different physical scenarios, and the choice of a specific foliation may introduce ambiguities in the interpretation of the data, potentially obscuring structure.

For instance, it is possible for the bifurcation to occur such that subsequent curvature peaks occur ever closer to the origin so that in the limit of infinite tuning, the curvature singularity forms at the origin. On the other hand, curvature singularities could form away from the origin. In either case, the blow-up could exhibit DSS or not. At the level of tuning currently possible in vacuum collapse, evidence suggests that there are families of initial data in which the singularity forms at the origin and others in which it does not. In either case, there is no evidence for strict DSS as in the spherical scalar field setting. But there are families that exhibit approximate DSS. Since the dynamics of the complex scalar field necessarily include those of the vacuum and real scalar field cases, all of this is highly relevant. Yet another possibility, which we do not presently have the tools to diagnose, is a true violation of weak cosmic censorship by the formation of a naked singularity on an open set of parameters $\left(p_{\textrm{min}},p_{\textrm{max}}\right)$ within a family.

To establish what happens in our setting, we first observe that with varying tuning to the threshold the curvature peaks appear at the same proper distance from the origin on the $z$-axis, at around~$z_p=0.0013$. Although this distance is foliation dependent, this suggests that the centers of collapse are distinct and are not expected to converge to a single point at the center of collapse for infinite tuning. To examine this in more detail we have constructed two special coordinate systems. In both, the idea is to build adapted coordinates to an assumed DSS centered at the origin. If the coordinates are indeed properly adapted to the symmetry in such a spacetime, dimensionless quantities will appear periodic in time over an extended region. Considering data over an entire coordinate patch constitutes a much more demanding test of self-similarity than looking along the world-line of a single observer at the origin.

\begin{figure*}[!t]
    \centering    
    \includegraphics[scale=0.825]{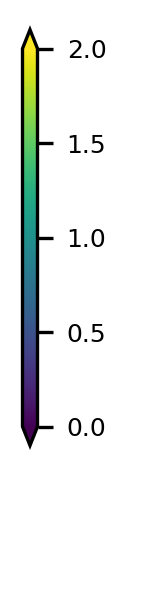} \,
    \includegraphics[width=0.635\columnwidth]{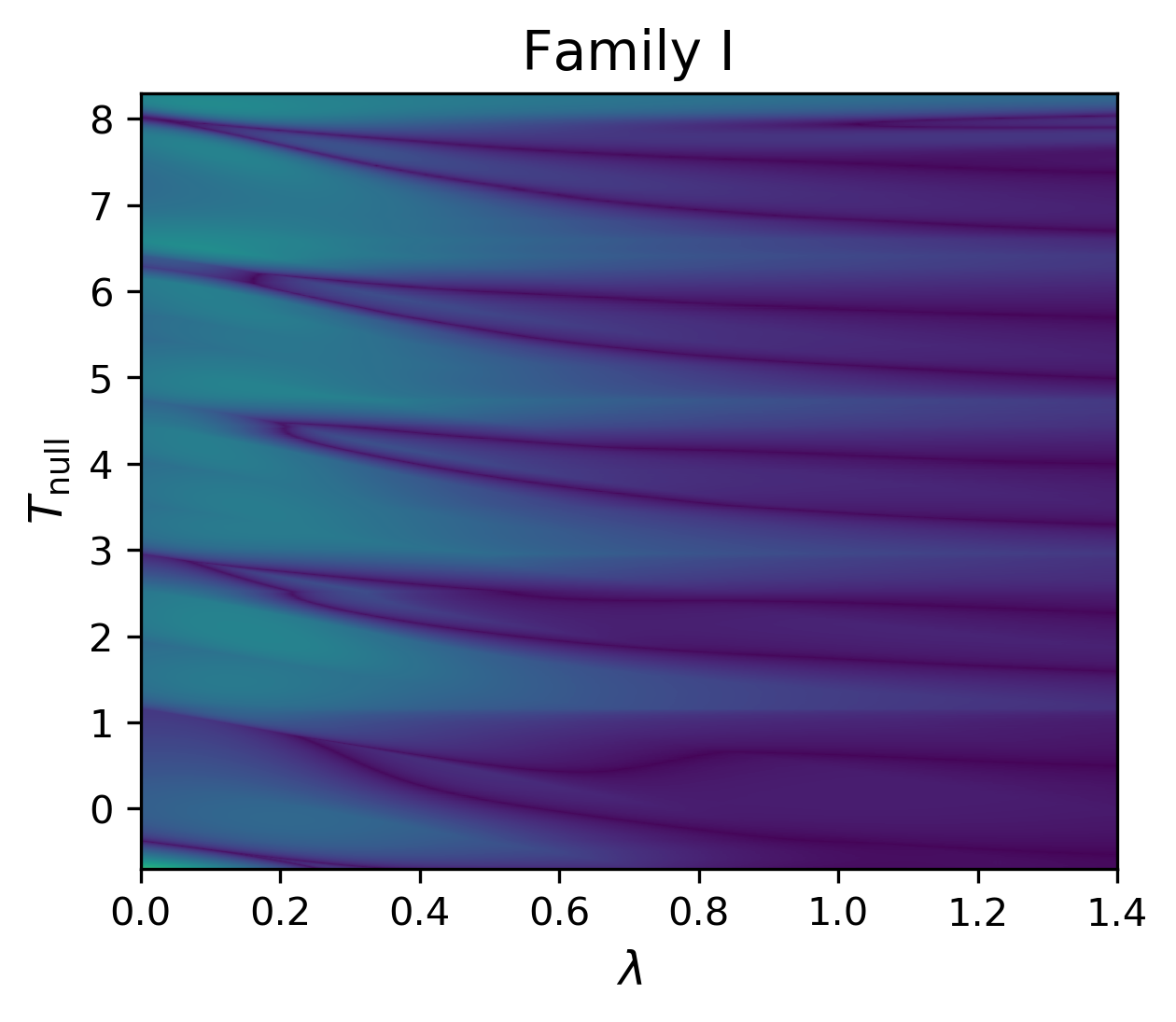}
    \includegraphics[width=0.635\columnwidth]{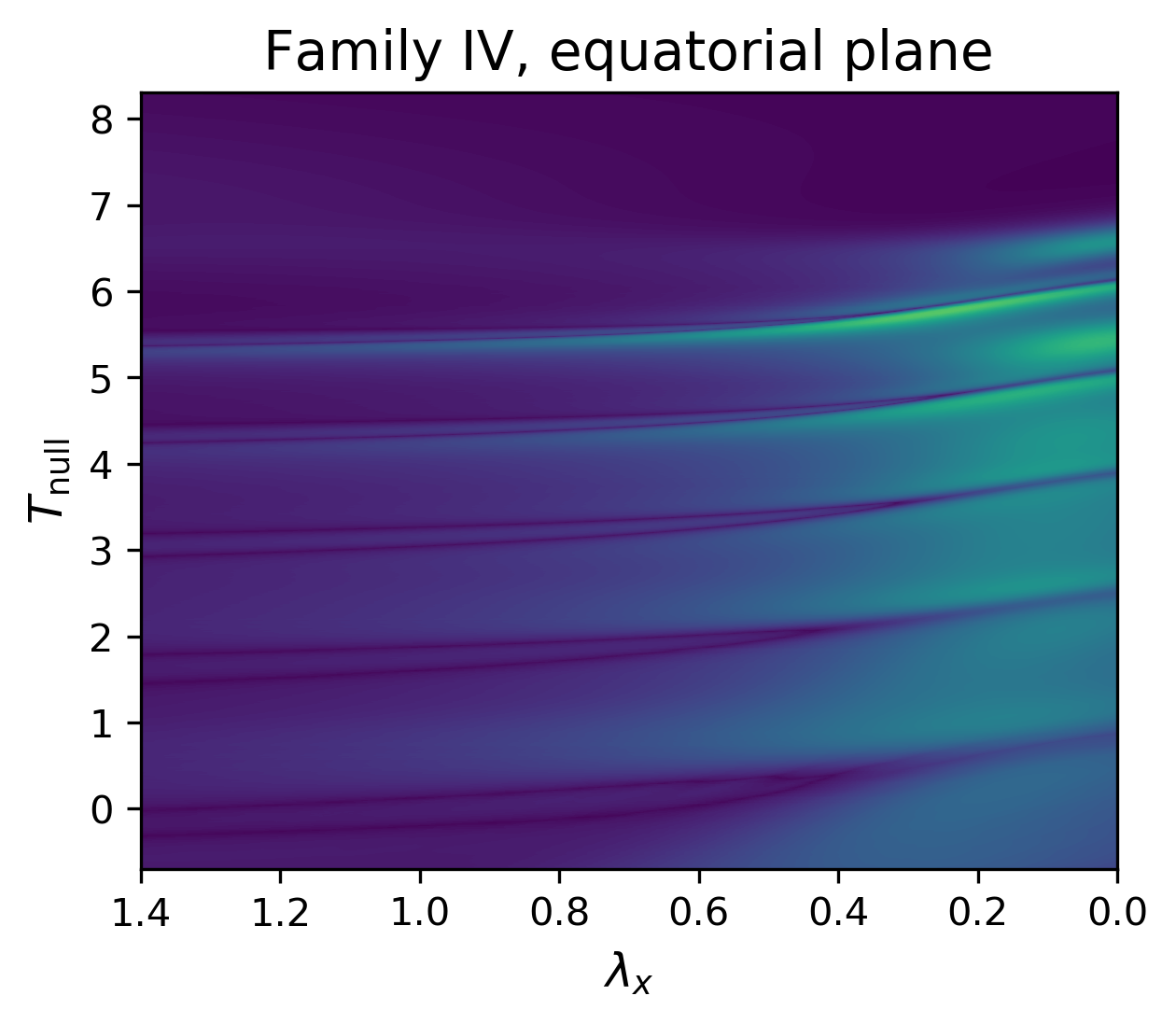}
    \includegraphics[width=0.59\columnwidth]{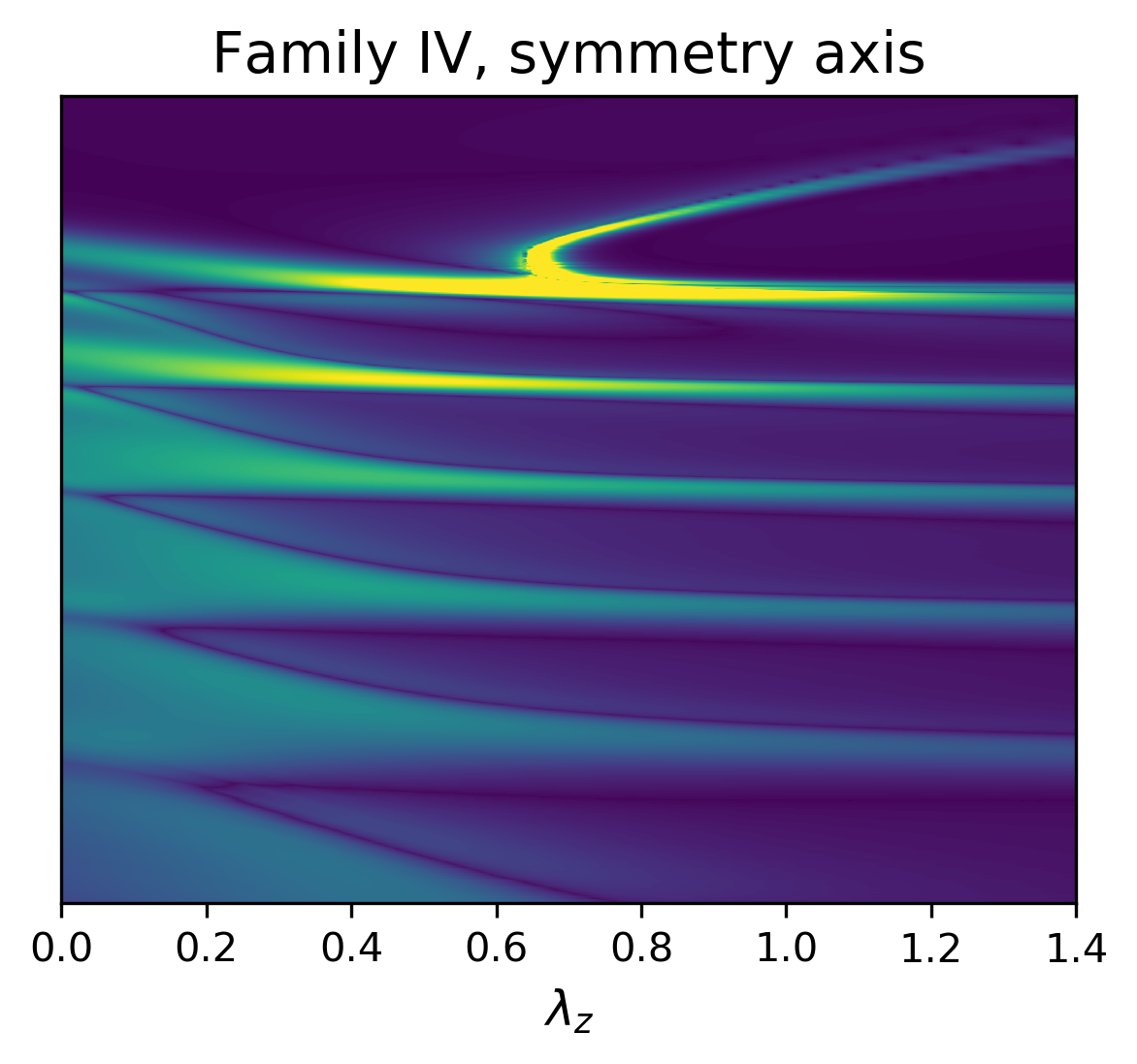}
    \caption{Heat maps of the trace of the normalized stress-energy tensor~$(\tau_*-\tau)|T_a{}^a|^{1/2}$  from two of our best-tuned spacetimes, plotted as a function of the single-null DSS-adapted coordinates constructed as described around Eq.~\eqref{eq:slow_time_null}. On the left the data from the Family I of Table~\ref{tbl:axi_families}, which are spherically symmetric, are plotted. As in Fig.~\ref{fig:rho_nullplots} periodicity here in~$T_{\textrm{null}}$ demonstrates the DSS nature of the spacetime. In contrast to the coordinates employed in Fig.~\ref{fig:properlength} we see that~$(T_{\textrm{null}},\lambda)$ do respect the symmetry of the spacetime. The center and right panel show data from Family IV, axisymmetric spacetimes with~$\epsilon^2=0.99$. In the center panel we integrated geodesics within the equatorial plane, and on the right on the symmetry axis. The asphericicity is clearly visible, as is a gradual thinning out of the features before the bifurcation of the center of collapse on the $z$-axis. The center and right plots do not display time-periodicity, disproving the hypothesis that the data are DSS with accumulation point at the center.}\label{fig:nullplots}
\end{figure*}

In our first attempt to build similarity coordinates, we work with the time coordinate~$T_p$, which is constructed exactly like~$T_{\textrm{null}}$ in Eq.~\eqref{eq:slow_time_null}, but now used to label leaves of the original foliation defined by our generalized harmonic coordinates. This is supplemented by the spatial coordinates~$X_p$ and~$Z_p$. By axisymmetry we need only consider these two directions. The first of these is constructed as
\begin{align}
    X_p =\frac{x_p}{|\tau_{*} -\tau|} \, , 
    \label{eq:Xp}
\end{align}
where~$x_p$ is the proper length from the origin along the~$x$-axis, computed as
\begin{align}
    x_p(x) =\int_{0}^{x} \sqrt{g_{xx}(t, x')} \, dx' \, . 
    \label{eq:proper_length}
\end{align}
The second is constructed similarly, but computing the proper distance from the origin along the symmetry axis (the~$z$-axis, in our conventions). In spherical symmetry of course we need only take one of these two. Our computations were performed using the gauge source functions~\eqref{eqn:DSS_compatible}, which pass only a necessary condition to adapt to DSS. In~\cite{cors2023formulation} it was found that in practice this choice does not result in DSS-adapted coordinates for the real scalar field. Given the results of Fig.~\ref{fig:rho_nullplots}, which suggests that in spherical symmetry the spacetime metric of the critical complex scalar field solution agrees exactly with the Choptuik solution, we expect the same outcome here. In Figure~\ref{fig:properlength}, we plot the strength of the complex scalar field of near-threshold spacetimes against these coordinates. In line with our expectation, away from the origin the plot does not look periodic in~$T_p$ even in the spherical case (left panel). The asphericity is clearly visible for family IV, $\epsilon^2=0.99$, in the middle and right panels. One sees in particular a bright spot on the~$z$-axis at~$Z_p\simeq0.6$. But since these coordinates fail to adapt to the DSS even in the spherical limit, they clearly cannot shed much light on the presence of symmetry in the axisymmetric setting.

Our second attempt to build similarity coordinates on the aspherical spacetimes is more informative. For this we follow exactly the procedure explained around Eq.~\eqref{eq:slow_time_null} for spherical spacetimes, but now for aspherical data we construct the affine parameter~$\lambda$ separately on the $x$-axis and on the symmetry axis. We have already seen in Fig.~\ref{fig:rho_nullplots} that this `re-slicing' procedure works well in the spherical setting, and so overcomes the issues with our foliation that are visible in Fig.~\ref{fig:properlength}. In Fig.~\ref{fig:nullplots} we plot the rescaled trace of the stress-energy tensor against these coordinates for the spherical (Family I, $\epsilon=0$) and most deformed (Family IV, $\epsilon=0.99$) initial data of Table~\ref{tbl:axi_families}. As expected, we find that the spherical data are cleanly periodic. We have plotted the data from families II and III in the same way, but do not show them in the figure. They give plots very similar to the spherical setup of Family I, with no sign of the bifurcation, albeit with a slightly deformed image in family III. Plotting instead the strength of the complex scalar field we again see clean periodicity, which demonstrates unambiguously that the lack thereof in the left panel of Figure~\ref{fig:properlength} is caused by the generalized harmonic foliation. Thus the single-null coordinates {\it are} adapted to DSS centered at the origin. The asphericity of family IV is once more clearly visible and, interestingly, the data are not periodic, disproving the hypothesis that the spacetime is DSS with accumulation point at the origin. We have not ruled out the possibility that a region of the spacetime is DSS with accumulation point elsewhere, but for now we do not have post-processing tools to construct similarity coordinates centered at an arbitrary point in spacetime.

\begin{figure}[!t]
    \centering
    \includegraphics[width=\columnwidth]{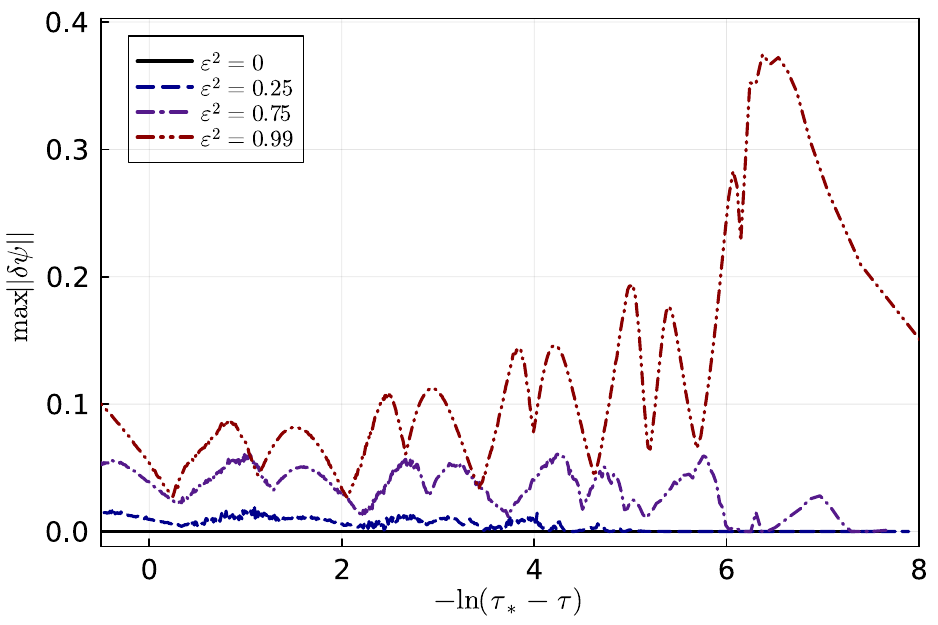}
    \caption{In this plot we show a quantitative diagnostic of the degree of asphericity of our data for literature comparison purposes. The quantity being plotted is the maximum difference between the values of the complex field amplitude on the $z-$axis and on the $x-$axis. More precisely, we identify the null coordinates on each axis, $\lambda_x$ and $\lambda_z$, and we take the differences at points that have the same value of that coordinate, $\lambda_x =\lambda_z$ . The plot reveals the maximum value of this difference against slow time and indicates how aspherical each dataset is at the level of initial data as well as at late times. The spherical family is, by definition, represented by the $x-$axis in this plot. It seems that the first two aspherical families present fluctuations of order~$0.01-0.05$ of the field difference between the  two axes, but converge to the spherical behavior at late times, where the threshold regime lies. The last aspherical family clearly presents a different pattern than the rest, since the degree of asphericity seems to grow with slow time and peak at $0.4$ after the bifurcation has occurred.}
    \label{fig:Baumgarte_Gundlach_comparison}
\end{figure}

Since the asphericity of the spacetime seems to be the key factor behind the departure from the spherical picture of critical collapse, it is desirable to have an unambiguous measure of the asphericity beyond the initial slice. Even if different codes evolve the same initial data, with~$\epsilon$ or~$s_z$ as a measure of initial asphericity, different numerical implementations and especially foliations might lead to different degrees of apparent asphericity in the time development, which might lead to apparent discrepancies between different studies. To compare with~\cite{choptuik2004critical,baumgarte2018aspherical} and the very recent single-null work of~\cite{Gundlach:2024mld,Gundlach:2024eds}, we assess the maximum difference of the scalar field amplitude at the poles and on the equatorial plane measured in our single-null DSS-adapted coordinates. The result is given in Fig.~\ref{fig:Baumgarte_Gundlach_comparison}. This is a good option because all published data can be evaluated in such coordinates to make a fair comparison. As we could expect from the fact that we only add an aspherical perturbation to the real part of the solution and not to the imaginary part, our~$\epsilon=0.75$ is indeed less aspherical than that investigated in~\cite{baumgarte2018aspherical}. Interestingly, our~$\epsilon=0.99$ family is more aspherical than the~$\epsilon=0.75$ data of~\cite{Gundlach:2024eds}, which could also account for the fact that no bifurcation is observed for that data. See~\cite{Gundlach:2024eds} for a detailed comparison and discussion of this.

The general solution space for the complex scalar field includes that of the real scalar field, which in turn includes that of vacuum. At the threshold of collapse, the complex scalar field therefore inherits all of the subtleties of both. A natural physical interpretation for our data then, already discussed in~\cite{Hilditch:2017dnw} for the real scalar field, is as a nonlinear admixture of scalar field and gravitational wave content. As we approach the threshold of collapse it could be that dynamics are driven primarily by either the scalar field or by gravitational waves, or that both play an important role. But the nature of gravitational waves in the nonlinear theory makes quantifying the relative strength of each complicated.

In a linear approximation against a background with enough symmetry one can identify a specific piece of the metric perturbation with a gravitational wave. When linearizing about the Minkowski spacetime for instance, we can take the transverse-traceless part of the metric perturbation. There is no such decomposition for the metric of a full solution to the nonlinear field equations. Therefore uncontroversial statements such as `there are no spherical gravitational waves', which make perfect sense in the linear theory, fail to hold in the full theory. Consider a vacuum solution in which a black hole forms and asymptotes at late times to the Schwarzschild solution. On physical grounds the solution is a gravitational wave throughout the collapse. Since there is no practical distinction between what are often labeled `waves' and the rest of the metric, it persists as such forever. From this point of view {\it any} solution to vacuum general relativity, including the Kerr spacetime, ought to be viewed as a pure gravitational wave. (The earlier statement should perhaps be amended to read, `there are no dynamical spherical gravitational waves'). General solutions, as in the case of the complex scalar field, are a combination of gravitational waves and matter.

A plausible physical interpretation for our spacetimes is that in the spherical~$\epsilon^2=0$ case collapse is driven primarily by the scalar field. But as~$\epsilon^2$ is increased, the~$l=2$ spherical harmonic content of the scalar field initial data triggers a gravitational wave. At sufficiently large~$\epsilon^2$ the gravitational wave may dominate the evolution and result in a bifurcation as also observed in vacuum collapse. We need observable quantities to evaluate this hypothesis. Ideally we could build a collection of non-negative spacetime scalars (foliation independent by definition) to measure the strength of the gravitational wave and scalar field content and, in such a way that a side-by-side comparison would be fair. Presently we do not have such a collection. In view of the discussion above any such scalar could not be constructed purely from the metric without derivatives. Likewise the connection is gauge dependent, so we turn to curvature and the stress-energy tensor, which is natural as a measure on the scalar field.

The Weyl tensor is defined as the trace-free part of the Riemann tensor. In 4 dimensions it is given by
\begin{align}
C_{abcd}&=R_{abcd}+\tfrac{1}{2}\left(R_{ad}g_{bc}-R_{ac}g_{bd}+R_{bc}g_{ad}-R_{bd}g_{ac}\right) \nonumber\\
&\quad+\tfrac{1}{6}R\left(g_{ac}g_{bd}-g_{ad}g_{bc}\right).
\end{align}
Since only traces of the Riemann tensor appear in the Einstein equations~\eqref{eqn:EEs}, the Weyl tensor is the part of the curvature that does not participate directly in the equations of motion. Motivated by these observations, it is common to associate the Weyl tensor with gravitational waves, and the Ricci tensor instead to matter content. An excellent example reinforcing this perspective is the Schwarzschild solution, for which the Weyl and Riemann tensors are identical. The Weyl tensor therefore serves as an obvious crutch to build scalars we seek. As a naive first option, we propose to quantify the relative strength of gravitational waves via the scalar
\begin{align}
W=C^{abcd}C_{abcd}\,,\label{eqn:W_Scalar}
\end{align}
as compared to the Kretschmann scalar,~$I=R^{abcd}R_{abcd}$, which includes also the matter contribution. Our physical interpretation above should imply that the bifurcation occurs when the contribution of the Weyl tensor to the Kretschmann scalar is comparable to, or dominates that of the traces.

\begin{figure}[!t]
    \centering
    \includegraphics[width=\columnwidth]{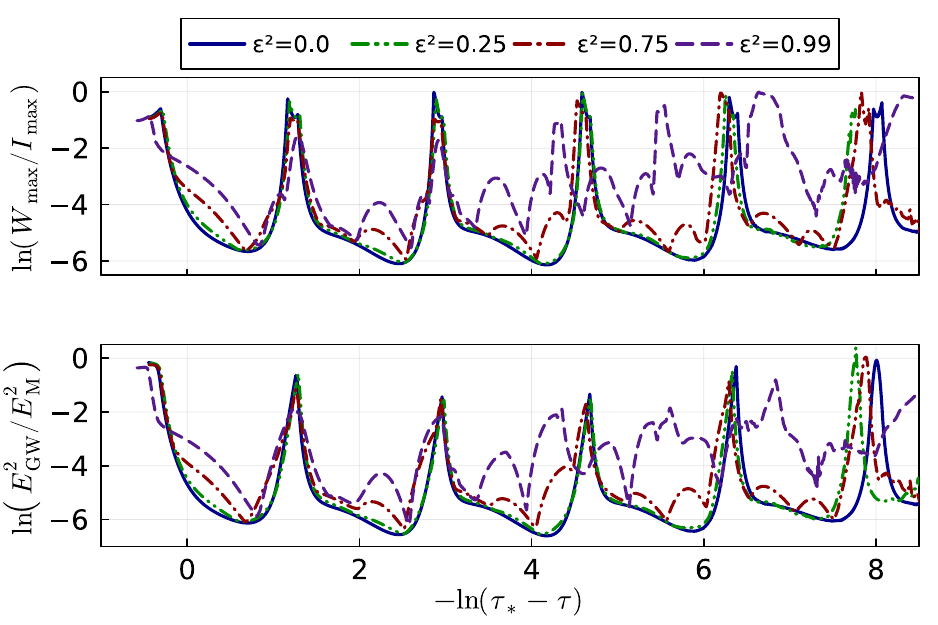}
    \caption{Top panel: logarithm of the ratio of the maximum of the scalar~$W$ and the Kretschmann scalar, $\ln(W/I)$, against similarity time, for the best-tuned members of each family (with fixed~$\epsilon^2$). We observe that as~$\epsilon^2$ increases, the contribution to the Kretschmann curvature scalar coming from the Weyl scalar begins to have a more complicated structure and progressively higher degree of significance. Bottom panel: logarithm of the ratio of the maximum of the energy scalars~$E_{\text{GW}}^2$ and~$E_{\text{M}}^2$, again as a function of similarity time. The results are qualitatively similar in both plots. See the main text for a more thorough discussion.
    }
    \label{fig:weyl}
\end{figure}

As an initial evaluation of the relative contribution to the total curvature throughout the evolution, in the top panel of Figure~\ref{fig:weyl} we plot the logarithm of the ratio of the scalar~$W$, defined by~\eqref{eqn:W_Scalar}, and the Kretschmann invariant as a function of slow-time defined by~\eqref{eq:slow_time_null}. This particular ratio is chosen, so that we plot dimensionless scalar quantities. For the families that start closer to spherical symmetry, the contribution of the traces of the Riemannn tensor dominates the collapse, except in brief periodic pulses in which the logarithm instantaneously vanishes, which means that the contributions are instantaneously equal. In these spacetimes it does appear that it is the scalar field driving the collapse. Yet we observe that as the degree of asphericity increases, scalar~$W$ does become more relevant. If one looks at the data of the location of the curvature maxima on the $z$-axis (which is not being plotted here), the bifurcation of the $\epsilon^2=0.99$ family starts being clearly visible at similarity time of around~$T_p\simeq5.47$. We observe also that, as seen in Figure~\ref{fig:weyl}, the $W$ dominates the dynamics around that time. Additionally, it is worth mentioning that there is an instantaneous peak of curvature off the origin at around~$T_p\simeq 4.3$ as well, which aligns with the peaks of the ratio of $W$ and the Kretschmann scalar. There is no sign whatsoever of bifurcation in the other families. In the most aspherical family, in which the bifurcation occurs, the curve follows a completely different oscillatory pattern from the beginning of the evolution, which we take as evidence supporting the physical interpretation given above. This interpretation additionally explains why the slightly aspherical families have a similar threshold solution to the Choptuik one, whereas the high asphericity ones display qualitatively and quantitatively different behavior.
\begin{figure*}[!t]
    \centering    
    \vspace{0cm}
    \includegraphics[scale=0.6]{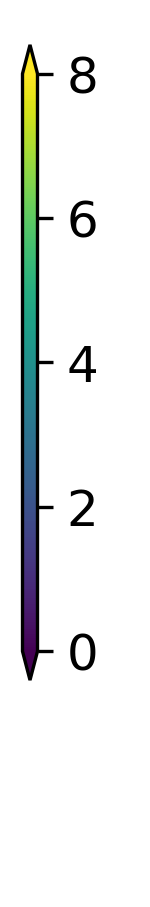} \,
    \includegraphics[scale=0.365]{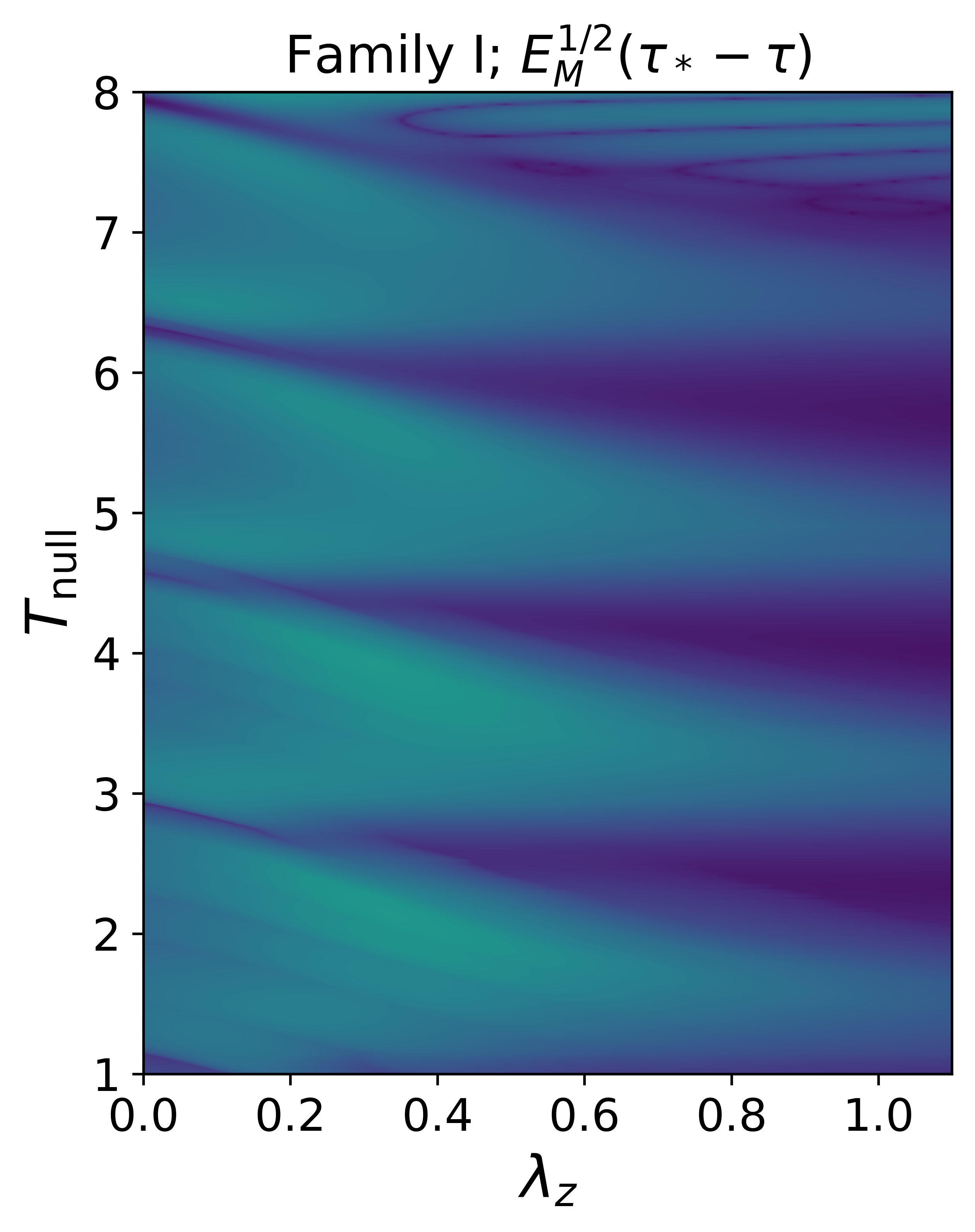} \,
    \includegraphics[scale=0.365]{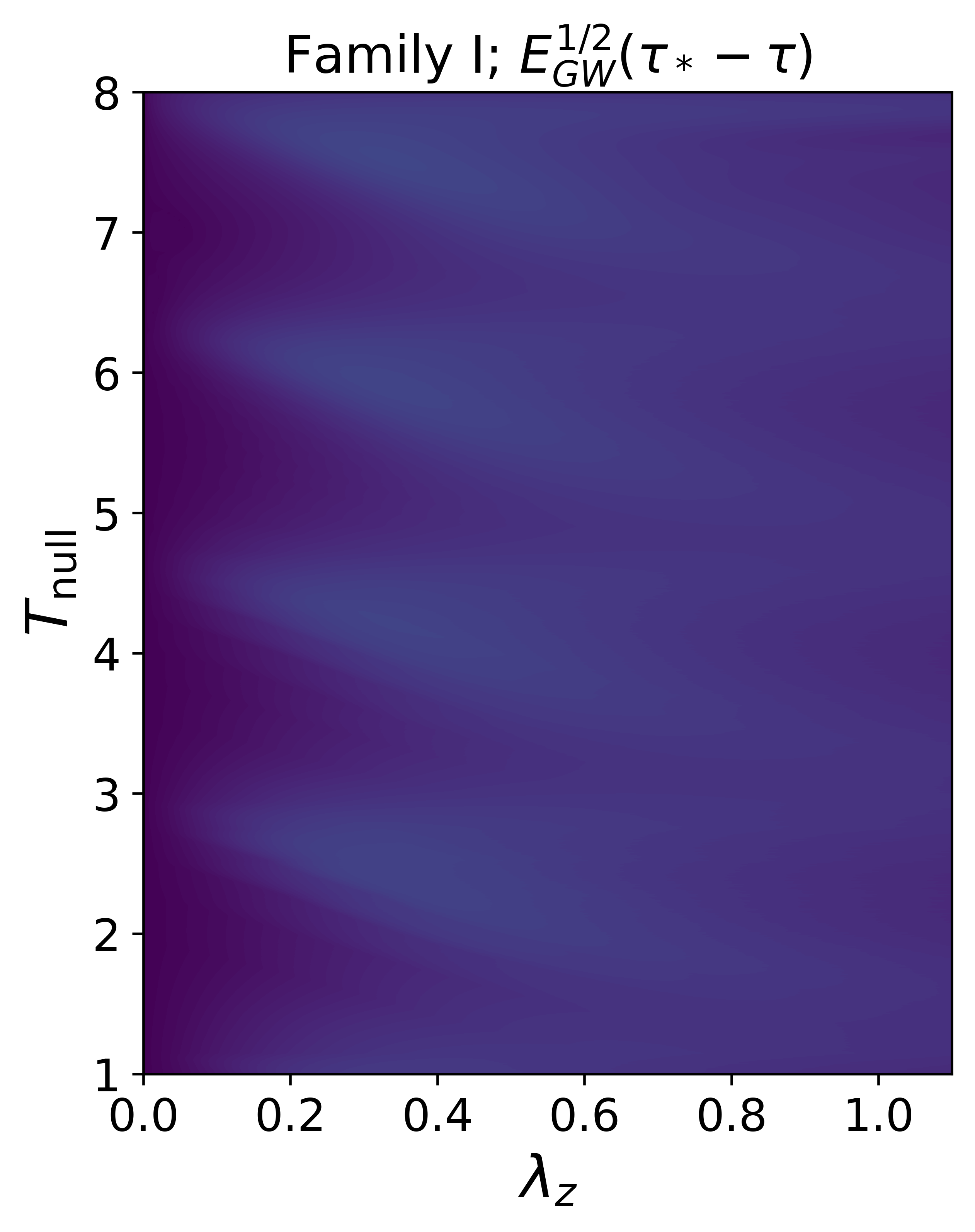} \,
    \includegraphics[scale=0.365]{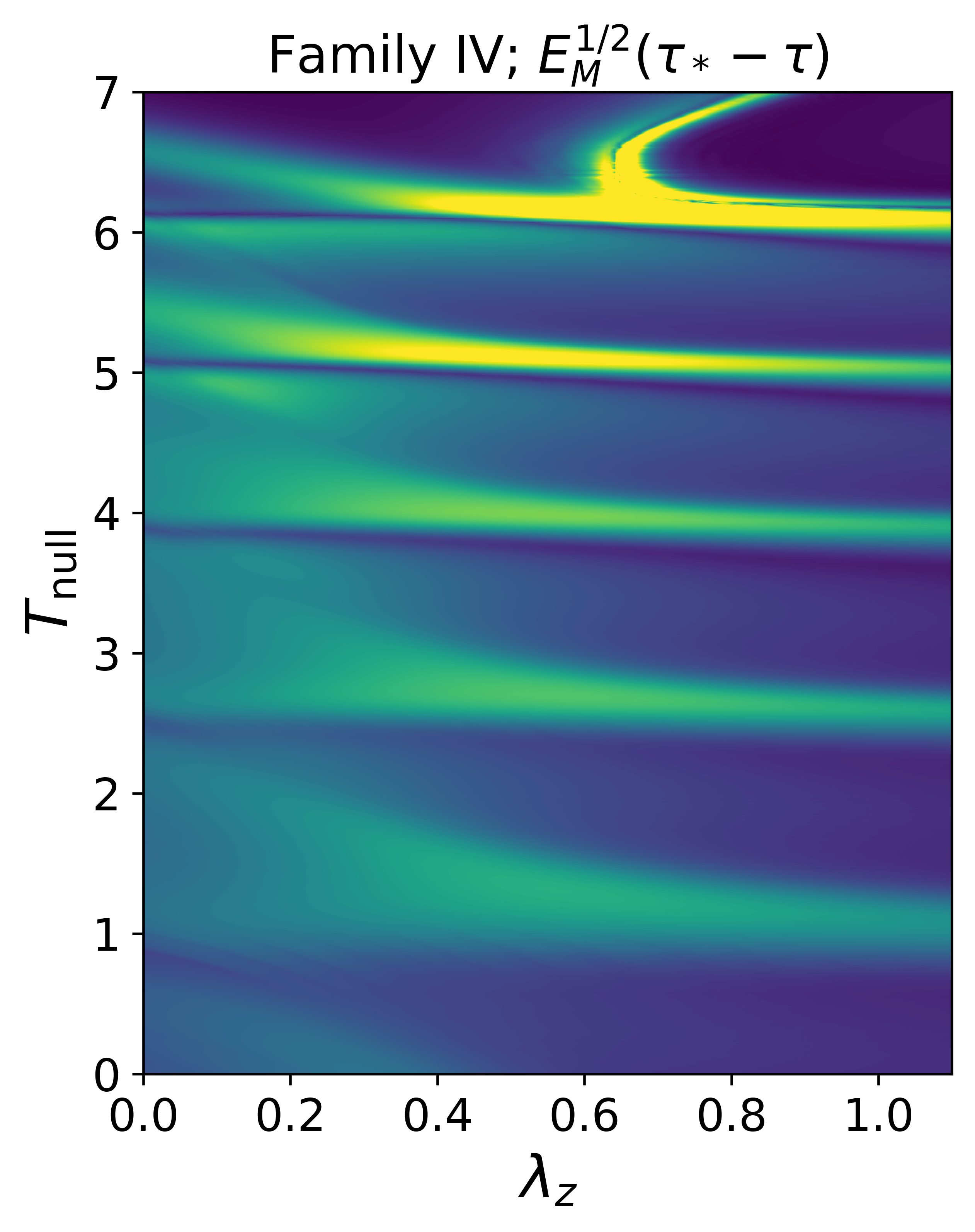} \,
    \includegraphics[scale=0.365]{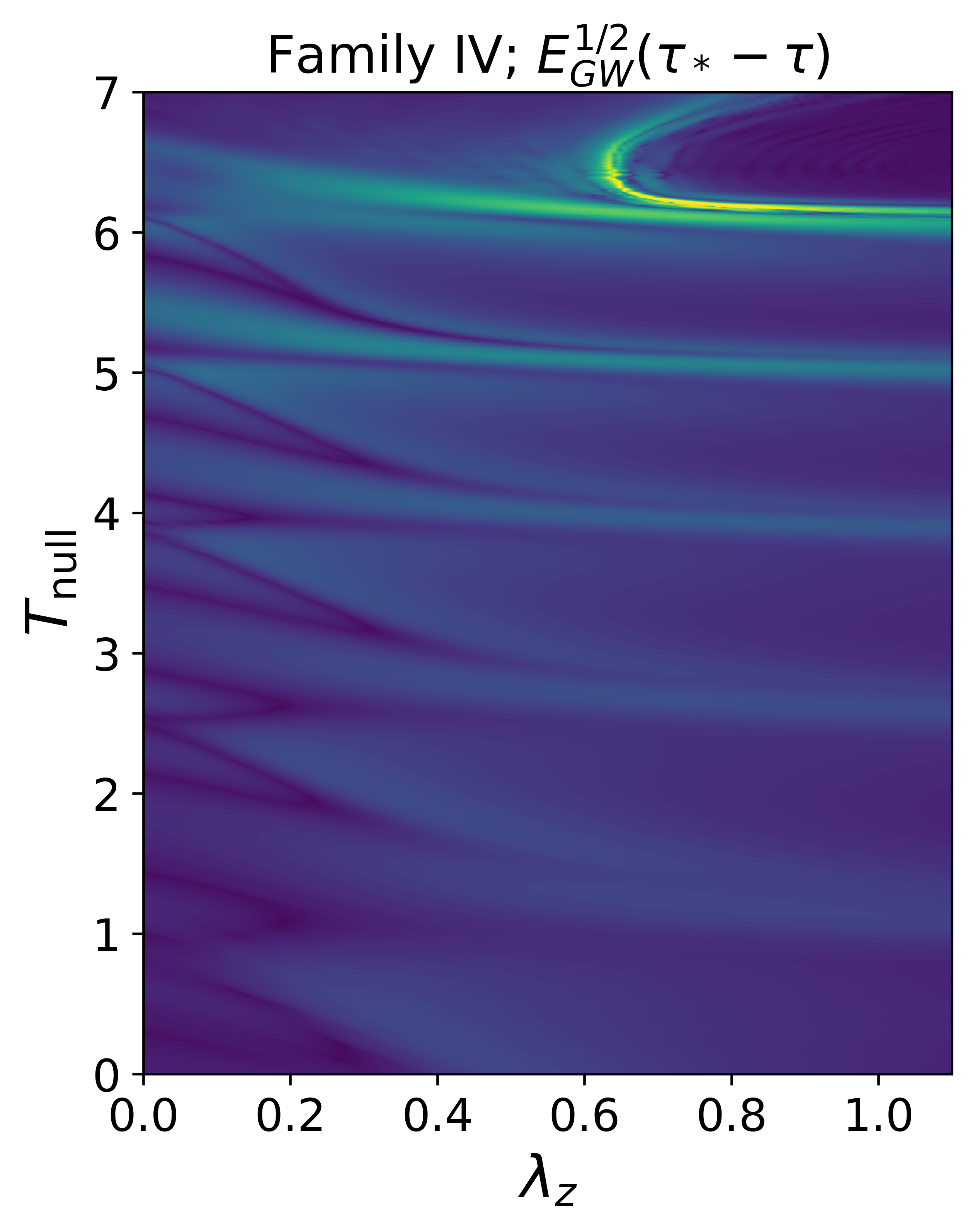} \,
    \caption{Color maps of the normalized energy scalars~$E_{\text{M}}^{1/2}(\tau_*-\tau)$ and~$E_{\text{GW}}^{1/2}(\tau_*-\tau)$ for our best tuned spherical (Family I) and most aspherical (Family IV) initial data setups, all along the symmetry axis in single-null similarity coordinates. As in Fig.~\ref{fig:nullplots}, thinning out of the features and the bifurcation is visible in the highly aspherical data. In the spherical case, shown in the left two panels, we see that relative to the matter terms, the instantaneous gravitational wave content remains small throughout, whereas in the aspherical spacetime, there is a small region where the two are comparable.} 
\label{fig:E_nullplots}
\end{figure*} 

The scalar~$W$ is manifestly foliation independent, but is not positive definite and so cannot unambiguously be interpreted as a measure of the strength of gravitational waves. Therefore we look for a supplementary quantity to consider. The Bel-Robinson tensor is constructed from the Weyl tensor according to
\begin{align}
T_{abcd}&=C_{aecf}C_{b}{}^{e}{}_{d}{}^{f}+{}^\star C_{aecf}{}^\star C_{b}{}^{e}{}_{d}{}^{f}\,,
\end{align}
with
\begin{align}
{}^\star C_{abcd}=\tfrac{1}{2}\epsilon_{ab}{}^{ef}C_{efcd} 
\end{align}
the Hodge dual of the Weyl tensor. The Bel-Robinson tensor is trace-free, symmetric on all indices and, in vacuum, divergence-free~$\nabla^aT_{abcd}=0$. It is constructed in a manner analogous to the way that the stress-energy tensor of electromagnetism is constructed from the Faraday tensor, and consequently has nice positivity properties; one obtains a non-negative scalar upon contraction with any future directed causal vector on all slots, and this quantity can be thought of as a density related to energy of gravitational waves. We now aim to construct a quantity that may be fairly compared with this. Given a tensor with the symmetries of the curvature~$W_{abcd}$, without assuming vanishing traces, we may define the symmetric tensor,
\begin{align}
Q(W)_{abcd}&= \left(W_{(a}{}^e{}_b{}^f W_{c}{}^g{}_{d)}{}^h
+{}^\star W_{(a}{}^e{}_b{}^f {}^\star W_{c}{}^g{}_{d)}{}^h\right) g_{eg} g_{fh}\,,
\end{align}
as a function of~$W_{abcd}$. Evidently this function agrees with the definition of the Bel-Robinson tensor when applied to the Weyl curvature~$Q(C)_{abcd}=T_{abcd}$. Upon contraction with a causal vector on all components, the object has interesting positivity properties regardless of the specific~$W_{abcd}$ tensor. We consider two interesting examples, and contract with~$n^a$, defined as above to be the future pointing unit normal to our foliation. We have,
\begin{align}    
E_{\text{GW}}^2=Q(C)_{nnnn}=T_{nnnn}=E_{ab}E^{ab}+B_{ab}B^{ab}\,,
\end{align}
with~$E_{ab}=C_{anbn}$ and~$B_{ab}={}^\star C_{anbn}$ the electric and magnetic parts of the Weyl tensor and indices~$n$ denoting contraction with the unit normal. We already gave a physical interpretation to this above. Second, we consider
\begin{align}
E_{\text{M}}^2&=Q(R-C)_{nnnn}\nonumber\\
&=\frac{16}{3}\pi^2(\rho+s)^2+32\pi^2 j_aj^a+16\pi^2 s^{\text{tf}}_{ab}s_{\text{tf}}^{ab}\,.
\end{align}
Here we have used the Einstein equations to replace contractions of the Weyl tensor, and have furthermore used the standard~$3+1$ expressions
\begin{align}
    \rho&=T_{nn}\,,\quad j_i=-T_{ni}\,,\nonumber\\
    s&=\gamma^{ij}T_{ij}\,,\quad s^{\text{tf}}_{ij}=T_{ij}-\tfrac{1}{3}\gamma_{ij} \gamma^{kl}T_{kl}\,,
\end{align}
for the decomposition of the stress-energy tensor. The latter renders the expression manifestly positive. We also observe that
\begin{align}    
0 \leq E^2_{\text{Total}}=Q(R)_{nnnn}\leq 2E_{\text{GW}}^2+2E_{\text{M}}^2\,.
\end{align}
We are not presently concerned with the differential properties of these objects, rather their interpretation as a local measure on the relative strength of gravitational wave and matter content, so we do not consider their time development. Comparing with the stress-energy tensor, we may think of the scalar~$W$ as being roughly analogous to the trace of the stress-energy tensor for matter; it is a true spacetime scalar, but can take negative values. Instead, like the energy density~$\rho$, the second set of scalars we have constructed here are foliation dependent but are positive and have units of energy density. Since neither satisfies simultaneously the points on our original wish-list for the collection of scalars, we use both to analyze our data.

In the lower panel of Fig.~\ref{fig:weyl} we plot the maximum, within each leaf of our foliation, of the logarithm of the ratio of~$E_M^2$ and~$E_{GW}^2$ as a function of~$-\ln(\tau_*-\tau)$ for our best-tuned member of each of the four families in Table~\ref{tbl:axi_families}. As in our comparison of~$W$ and~$I$ in the top panel, we see that substantial deviations from the periodic structure appear as the asphericity increases. We also see that the gravitational wave energy scalar grows substantially with asphericity. In Fig.~\ref{fig:E_nullplots} we plot heat maps for our best tuned data for Families~I and~IV, our spherical and most aspherical setups, respectively. Specifically,  we compare the energy scalars~$E_{\text{M}}^{1/2}(\tau_*-\tau)$ and~$E_{\text{GW}}^{1/2}(\tau_*-\tau)$ in the single null DSS-adapted coordinates used also in Fig.~\ref{fig:nullplots}. As in Fig.~\ref{fig:nullplots} the bifurcation is clear. We furthermore observe that, whereas in the spherical data the gravitational wave energy remains far smaller than the scalar field terms, in the aspherical data the relative size of the two scalars becomes comparable as the evolution develops. Interestingly, although the features in the aspherical data clearly do not agree with the spherical solution, structure does remain.

\begin{figure}[t]
    \centering
    \includegraphics[width=\columnwidth]{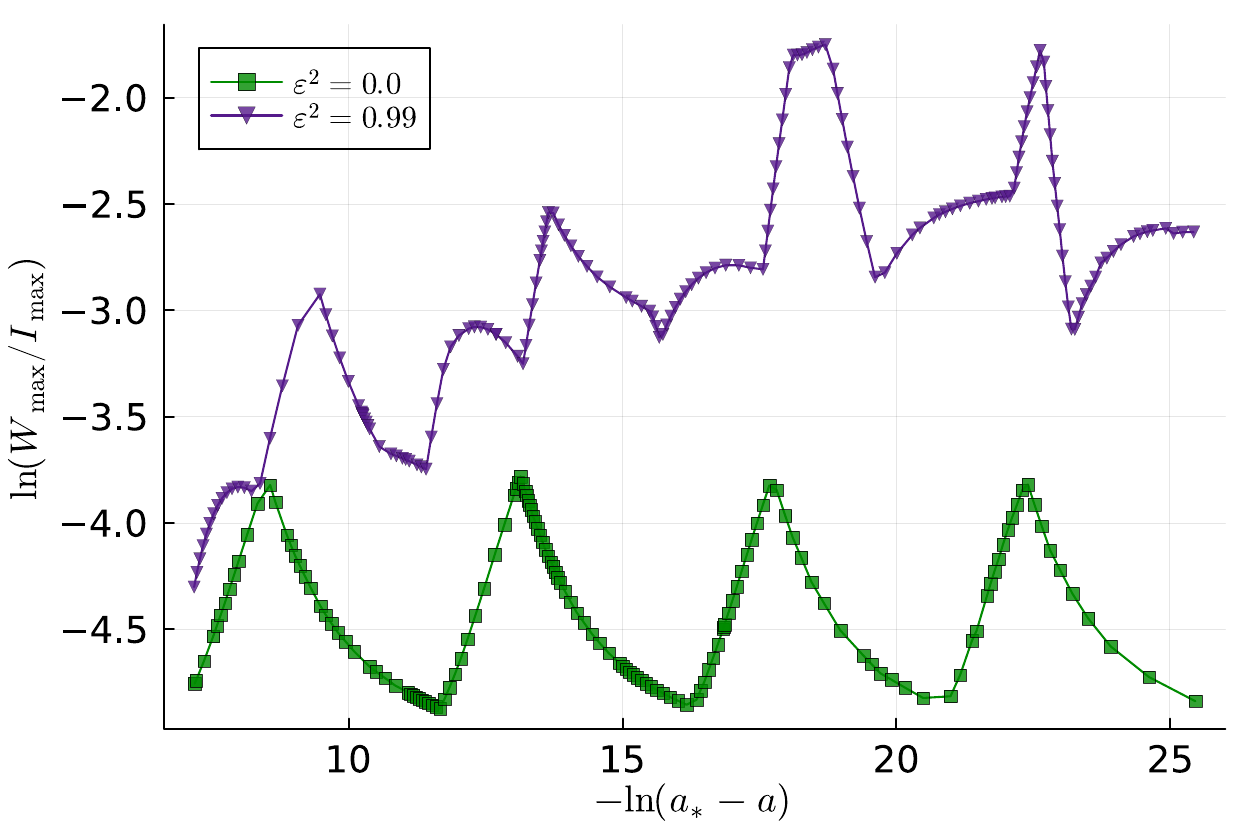}
    \caption{  
    Parametric plot of~$\ln(W_{\textrm{max}}/I_{\textrm{max}})$, against the distance in phase space from the threshold, using the global maxima over~$W$ and~$I$ from each of the evolutions. In this plot we focus just on the two extreme families of our searches, namely the spherical one with~$\epsilon^2=0.0$, and the highly aspherical~$\epsilon^2 = 0.99$ case. Since the Weyl and Kretschmann invariants have the same units, in the standard picture of critical collapse they scale in the same way near the threshold. Their ratio is then expected to reveal a universal DSS structure in case universality holds. Instead, to the extent that we can tune, we observe, in a fully foliation independent way, that in the~$\epsilon^2=0.99$ family there is no evidence either that the limiting solution agrees with the Choptuik spacetime, or indeed that it is DSS.}
    \label{fig:fol_indep_comparison}
\end{figure}

Taken together, Figs.~\ref{fig:nullplots} and~\ref{fig:E_nullplots} indicate that near the threshold, our most aspherical family does not exhibit discrete self-similarity centered at the origin and is thus not a single copy of the Choptuik solution. That said, we could still reasonably describe a portion of the spacetime as approximately DSS about the center. They also show that the gravitational wave component of the spacetime becomes much more significant in the highly aspherical regime. Figure~\ref{fig:weyl} corroborates this, and furthermore suggests that a lack of DSS may be more clearly diagnosed by looking at a more complete set of scalar fields.

Yet obvious questions remain. Could the spacetime be DSS with the accumulation point centered away from the origin? If so, are the bifurcated centers copies of the Choptuik solution? Are the dynamics instead becoming close to those in the vacuum setting?

As mentioned above, we do not have tools to search for and construct, or indeed to directly rule out the existence of similarity coordinates centered at an arbitrary spacetime event. We therefore need coordinate independent means to try and answer the first question. The obvious strategy is to study scalars as a function of the parameter~$a$. When the usual DSS scenario is realized, one expects that the maxima of all scalars with units of length would display the same power law growth and periodic wiggle as a function of the logarithmic distance to the threshold, see~\cite{Gundlach:1996eg,Hod:1996az,Garfinkle:1998va}. In addition to Fig.~\ref{fig:axiscalingwl}, we therefore fit a power law through the maxima of the spacetime scalars~$I$ and~$W$ and consider their ratio, looking for a universal periodic function. In the spherical case we find that we can fit the same power law through both curves, with~$\gamma_I=\gamma_W=0.38\pm0.01$ immediately recognizable from the Choptuik solution. In Family IV instead we find~$\gamma_I=0.33\pm0.01$ and~$\gamma_W=0.35\pm0.01$. Again, the errors quoted here are just from regression, so one has to be careful not to overinterpret the difference. In figure~\ref{fig:fol_indep_comparison} we plot the ratio of the maximum of~$\ln(W/I)$ as a function of~$-\ln(a_*-a)$. The curves from the spherical and aspherical families do not agree. Once more, the evidence suggests that gravitational wave content, measured here by~$W$, is substantially higher in the highly aspherical family than in the spherical counterpart. The difference that we find in the power-law fits is visible as a very slight drift in the~$\epsilon=0.99$ curve. By far the most important result of the plot, however, is that whilst there is a clear periodicity in the curve from the spherical data, this structure is completely absent in the aspherical family. The fact that the two curves do not agree suggests that the leading dynamics in the limiting spacetimes do not either, and, thus, the bifurcated region does not appear to be a copy of the Choptuik solution. The fact that the aspherical family does not return a periodic function suggests that the limiting spacetime is not DSS either. Although both must be taken under the global caveat that any numerical approach can tune only to a limited extent, these findings are compatible with those of Fig.~\ref{fig:axiscalingwl}. To verify this directly we have rendered a simple colormap of the ratio~$W/I$ from our best-tuned spherical and most aspherical spacetimes. Structurally, these plots closely resemble the first and third panels of Fig.~\ref{fig:E_nullplots} so we do not include them. Visually they show that the region surrounding the bifurcated center of collapse in the most aspherical data is not a straightforward transformation of the Choptuik solution in similarity adapted coordinates. It is very interesting that in~\cite{Reid:2023jmr} there is an example of a different aspherical family, for the real scalar field, which bifurcates but {\it does} reveal two copies of the Choptuik solution, so there is much more to be understood about the solution space. It would therefore be good to have more families and better tuned data to investigate all of this in more detail.

Finally, to compare with the vacuum threshold, we have rendered plots like those of~\cite{Ledvinka:2021rve,SuarezFernandez:2020wqv}, which showed repeated features. This is done by plotting scalar functions along distinct integral curves of~$n^a$ passing through subsequent localized peaks in the curvature. Perhaps a better procedure would be to plot along geodesic curves through the peaks, but we leave this for future work. The curves are parameterized by proper time, off-set for each individual curve so that~$\tau=0$ at the peak, and then scaled according to the curvature scale at the peak (see the references above for details). In the spherical setting we obtain curves similar to those in Fig.~\ref{fig:Realvscomplex}, but for single echoes at a time. As one would expect, each of these curves agrees well with the next. Doing the same in our best-tuned, most  aspherical data, there is tentative evidence for repeated features forming away from the origin. Quantities associated with the scalar field, such as the amplitude of the scalar field itself, or the trace of the stress-energy tensor, display these features most convincingly. But these features do not agree closely with those of the spherical data. Furthermore, certain features do appear similar to earlier pure vacuum data. We already know from the discussion above that the threshold spacetime for this family is not close to the Choptuik solution. Overall, these plots may suggest that whilst the gravitational wave content contributes significantly to the dynamics of the spacetime, the threshold spacetime is quite far from a pure gravitational wave collapse as well, an interpretation compatible with all of the results above. For example the maximum of the Kretschmann scalar, which contains both gravitational wave and matter contributions, is an order of magnitude larger than the scalar~$W$, which we associate with gravitational waves. What is more, whilst there is a small upward drift in the data of Fig.~\ref{fig:fol_indep_comparison}, we cannot be confident that~$W$ will begin to dominate with greater tuning, since the latter part of the curve does appear to flatten. Given the uncertain nature of the threshold of collapse in vacuum, and the challenge of interpretation of these latter plots, which reveal at most two such features per quantity, we suppress them until a more satisfactory analysis can be given.

\section{Conclusions}
\label{sec:conclusions}

We explored for the first time twist-free axisymmetric spacetimes with a massless complex scalar field near the threshold of gravitational collapse. Our aim was to contribute to the literature regarding this specific model, in particular to examine aspherical perturbations of spherical threshold solutions. To do so we used the pseudospectral numerical relativity code~\textsc{bamps} to calculate solutions, and bisected toward the threshold. Recent developments of adaptive mesh-refinement~\cite{renkhoff2023adaptive}, improved constraint damping~\cite{cors2023formulation} and our complex scalar field implementation~\cite{Atteneder:2023pge} formed a crucial foundation for the investigation. We have carefully tuned to the threshold, maintaining constraint violations five or more orders of magnitude smaller than curvature invariants of interest. Our results are divided between two basic configurations.

Our first set of findings concerns spherical symmetry configurations, for which we tuned four distinct families of initial data to the threshold of collapse. Although closely related models have been studied before, as far as we are aware, this is the first time that the spherical configuration for the massless complex field has been carefully treated. We found strong evidence that there is a universal critical spacetime metric which agrees with that of the real scalar field. This metric is characterized partially by the echoing period~$\Delta\simeq3.4$. Perturbations from the critical solution lead to power-law growth of scalar quantities with exponent~$\gamma\simeq0.37$ also familiar from the spherical real scalar field~\cite{choptuik1993universality}. Since the complex scalar solution space includes that of the real field, this was certainly a possibility a priori, but requires the real and imaginary parts of the field to concoct precisely the stress-energy tensor of the Choptuik solution, and so was a surprise to us. Interestingly similar behavior in which, in a certain limit, the dynamics of a more complicated system agree with those of a simpler model, occurs at the threshold of blow-up for several of the models examined in~\cite{SuarezFernandez:2020wqv}. This surprise is heightened by the fact that, in the general axisymmetric setting with twist (angular momentum)~\cite{choptuik2004critical} the threshold of collapse for real and complex fields certainly do not coincide; in that context the real field cannot support angular momentum whereas the complex field can.

The second set of findings concerns aspherical configurations,
for which we find a diminishing tendency in the period of the central absolute value of the complex scalar field and the scaling law power exponents with respect to increasing asphericity. In highly aspherical simulations we do not observe periodicity in quantities expected to exhibit periodic behavior according to the standard picture of critical collapse. We observe this lack of periodicity both within well-tuned individual spacetimes, and parametrically as a function of the distance to the threshold of collapse. In qualitative agreement with results for the real scalar field~\cite{choptuik2003critical,baumgarte2018aspherical}, we find the bifurcation of the center of collapse for the complex field on to the symmetry axis, but away from the origin. Plotting spacetime scalars in coordinates adapted to an assumed self-similarity revealed that our very aspherical data are not DSS with accumulation point at the origin. Because they lie in a separate symmetry class, these threshold solutions also cannot agree with those of~\cite{choptuik2003critical}, which have non-vanishing angular momentum. Observing that as the~$l=2$ harmonic in the scalar field is increased in the initial data, our results begin to resemble vacuum collapse results, we attempted to give a physical interpretation as a competition between matter and gravitational wave content. To support this we considered curvature scalars built from the Riemann and Weyl tensors and found that the latter, associated with gravitational waves, indeed becomes comparable to the former as the level of asphericity increases.

Technical challenges to our numerical work and data analysis remain. The most obvious pertains to our classification scheme. If data can be evolved long enough to observe dispersion, we can confidently classify data as subcritical. But as the level of asphericity increases, the search for apparent horizons becomes difficult. At the greatest level of asphericity, in our most tuned data, we had to resort to weaker diagnostics, namely explosion of curvature scalars with constraint violation remaining small and both appearing smooth. Although our gauge choice does not adapt to DSS even in the spherical setting, in contrast to earlier numerical work~\cite{Hilditch:2017dnw}, to the degree of tuning we have managed here, it does not appear to suffer from coordinate singularities. The good behavior of the constraints and gauge is largely due to the improvements identified in~\cite{cors2023formulation}. Our difficulties in finding apparent horizons are familiar from the vacuum case. We suppose that they are caused by the star-shape assumption in our apparent horizon finder. Therefore a near-term goal is to implement the suggestions of~\cite{Pook-Kolb:2018igu,Poo20} which do away with this assumption. More generally, it would be good to have more sophisticated tools for analysis, as presently there is the danger that we overlook structure. We can, for instance, now readily construct coordinates adapted to DSS, and thus to the nascent singularity, with accumulation point at the origin. One could envisage automating a procedure to do so with the blow-up occurring elsewhere. But for now we have no such algorithm. Likewise, we suspect that more careful thought could reveal additional diagnostics that could be informative. With such tools we may eventually be able to identify a physical mechanism for bifurcation as gravitational wave content takes away.

There are a large number of potential avenues for follow-up work. The most obvious question involves the functional form of the dependence of~$\Delta$ and~$\gamma$ on the degree of (twist-free, axisymmetric) asphericity. This needs to be addressed both for the real and complex scalar field, and `just' requires us to take dense values of~$\epsilon^2$, which we have so far sampled only very sparsely. This should be done for prolate (positive $\epsilon^2$) and oblate (negative $\epsilon^2$) deformations and with initial data containing various different harmonics. The answer to this important question would indicate whether or not the spherical threshold solution is indeed isolated in solution space from other threshold solutions, and would rule out any disagreement with~\cite{Reid:2023jmr}, in which aspherical initial data were treated but where no drift of the critical parameters was observed. Their computations were made with pure~$l=0,1$ initial data for the real scalar field, whereas in the present study we posed a combination of~$l=0,2$ for the complex field, so although there is a qualitative difference, there is no explicit disagreement. It will also be important to consider the threshold of collapse of the complete axisymmetric solution space with twist for the real and complex fields, where qualitative differences must manifest. In the presence of angular momentum, we expect to recover the results of~\cite{choptuik2004critical}. It remains to perform very well tuned simulations for the real and complex scalar field models with absolutely no symmetry.

Beyond early foundational work on the spherical scalar field~\cite{Christodoulou:1986zr,Christodoulou:1991yfa,Christodoulou:1994hg}, to which the relationship with numerical evolutions was self-evident, there is an array of recent rigorous mathematical work on gravitational collapse. For instance, there are now various proofs on the formation of trapped surfaces without symmetry. The initial breakthrough of~\cite{Chr08}, for instance, has been extended to include small trapped surfaces~\cite{An:2019ybm}, scalar field matter~\cite{Zhao:2023czp}, electromagnetic~\cite{An:2020doc} and Yang-Mills~\cite{Athanasiou:2023ncp} fields. There are existence results for naked singurities in vacuum~\cite{Rodnianski:2017cvt,Rodnianski:2017cvt,Rodnianski:2019ylb,Shlapentokh-Rothman:2022byc,Shlapentokh-Rothman:2022uji}, and even a demonstration that naked singularities in the scalar field model will become trapped when subjected to a 3d perturbation~\cite{An:2024jqn}. Differences in gauge, coordinates, and indeed the entire mathematical framework employed have resulted in a gap between such mathematical and numerical work. Beyond these obvious issues, which may simply require effort to translate, there may be physical differences. It is interesting, for example, that despite the unavoidable complication of these proofs, they rely on a simple physical mechanism to drive collapse. In contrast, at least naively, as we tune numerically to the threshold of collapse in aspherical spacetimes, the mechanism for black hole formation appears anything but simple. An important goal for the future is therefore to understand and eliminate this divide.

From a purely computational point of view, the simplest follow-up would be the inclusion of a linear or a non-linear potential in the matter sector of the action, in order to allow boson stars as final states of the evolution and let us explore a wider range of collapse scenarios. This is already implemented and would potentially let us make contact with our studies of boson stars~\cite{Atteneder:2023pge}, but more importantly generalize the work of~\cite{jimenez2022critical} on the massive complex scalar field beyond spherical symmetry. Both of these possible changes enrich the solution space yet further, so it might instead be desirable to retreat to a simpler model. Although scalar field models were the first to be investigated historically, probably due to their mathematical simplicity, the Choptuik solution is difficult to treat. Physically speaking, the large scaling period necessitates~$e^\Delta\approx31$ times more resolution to accurately resolve each subsequent echo. In coordinates poorly adapted to the spacetime structure, yet more resolution could be required. In double-precision arithmetic, combined with the power-law parameter~$\gamma\simeq0.37$ this effectively limits the maximum number of echoes we can uncover to three~\cite{Gundlach:2019wnk}. It is desirable to study aspherical perturbations of spherical critical solutions from a matter model with a smaller scale period. We expect qualitatively similar findings from such a model, but with the advantage that measurements of perturbations to the period would be possible much more accurately with more repetitions to work with. An excellent candidate for this is the Yang-Mills field, whose spherical critical solution manifests with~$\Delta\simeq 0.74$ (see~\cite{Choptuik:1996yg,Maliborski:2017jyf} for details).

In closing, our results indicate that near the threshold of gravitational collapse, in the presence of sufficiently large aspherical perturbations, massless complex scalar field spacetimes start to exhibit complicated behavior reminiscent of that observed in vacuum collapse. Improving our understanding of the threshold of vacuum collapse will therefore be crucial even beyond its own obvious fundamental importance. This phenomenology is furthermore in qualitative agreement with behavior of the real scalar field, and is likely indicative of a broader trend, with many more models yet to be explored in axisymmetry and beyond.

\section*{Acknowledgments}

We are grateful to B. Br\"ugmann, T. Giannakopoulos, C. Gundlach, A. Khirnov, D. Nitzschke, R. Pinto Santos, U. Sperhake, A. Vañó-Viñuales and M. Zilh\~{a}o for helpful discussions and feedback on various aspects of the work. 

We acknowledge financial support provided by FCT/Portugal through grants 2022.01324.PTDC, PTDC/FIS-AST/7002/2020, UIDB/00099/2020 and UIDB/04459/2020. D.C. acknowledges support from the by the STFC Research Grant No. ST/V005669/1 ``Probing Fundamental Physics with Gravitational-Wave Observations''. D.C. and F.A. were supported in part by the Deutsche Forschungsgemeinschaft (DFG) under Grant No. 406116891 within the Research Training Group RTG 2522/1. H.R.R acknowledges support from the Funda\c c\~ao para a Ci\^encia e Tecnologia (FCT) within the projects UID/04564/2021, UIDB/04564/2020, UIDP/04564/2020 and EXPL/FIS-AST/0735/2021 and the H2020 ERC Advanced Grant ``Black holes: gravitational engines of discovery'' grant agreement no. Gravitas–101052587. The figures in this article were produced with~\textsc{Plots.jl}~\cite{christ2022plots}, \textsc{Matplotlib}~\cite{Hunter:2007}, A. Khirnov's~\textsc{nr\_analysis\_axi} package~\cite{Khirnov:nranaaxi} and \textsc{ParaView}~\cite{ayachit2015paraview,ahrens200536}. Several calculations were performed using the~xAct package~\cite{xAct} for Mathematica. Numerical simulations were performed at the Leibniz Supercomputing Centre (LRZ), supported by the project~\textsc{pn36je}. The authors thankfully acknowledge the computer resources, technical expertise and assistance provided by CENTRA/IST. Computations were performed at the cluster ``Baltasar-Sete-Sóis'' and supported by the H2020 ERC Consolidator Grant ``Matter and strong field gravity: New frontiers in Einstein's theory'' grant agreement no. MaGRaTh-646597.

\bibliography{main.bbl}

\end{document}